\documentclass[useAMS,usenatbib]{mn2e}
\usepackage{color}
\usepackage{amssymb,amsmath}
\usepackage{graphicx}

\title[Parametrising arbitrary galaxy morphologies: potentials and pitfalls]{Parametrising arbitrary galaxy morphologies: potentials and pitfalls}

\author[R. Andrae, K. Jahnke \& P. Melchior (2010)]{Ren\'e Andrae$^{1}$\thanks{E-mail:
andrae@mpia-hd.mpg.de}, Knud Jahnke$^{1}$ and Peter Melchior$^{2}$\\
$^{1}$Max-Planck-Institut f\"ur Astronomie, K\"onigstuhl 17, 69117 Heidelberg, Germany\\
$^{2}$Institut f\"ur Theoretische Astrophysik, Zentrum f\"ur Astronomie, Albert-Ueberle-Str. 2, 69120 Heidelberg, Germany}

\begin{document}

\date{Accepted 2010 September 10. Received 2010 April 1.}

\pagerange{\pageref{firstpage}--\pageref{lastpage}} \pubyear{2010}

\maketitle

\label{firstpage}

\begin{abstract}
Given the enormous galaxy databases of modern sky surveys, parametrising galaxy morphologies is a very challenging task due to the huge number and variety of objects. We assess the different problems faced by existing parametrisation schemes (CAS, Gini, $M_{20}$, S\'ersic profile, shapelets) in an attempt to understand why parametrisation is so difficult and in order to suggest improvements for future parametrisation schemes.

We demonstrate that morphological observables (e.g. steepness of the radial light profile, ellipticity, asymmetry) are intertwined and cannot be measured independently of each other. We present strong arguments in favour of model-based parametrisation schemes, namely reliability assessment, disentanglement of morphological observables, and PSF modelling. Furthermore, we demonstrate that estimates of the concentration and S\'ersic index obtained from the Zurich Structure \& Morphology catalogue are in excellent agreement with theoretical predictions. We also demonstrate that the incautious use of the concentration index for classification purposes can cause a severe loss of the discriminative information contained in a given data sample. Moreover, we show that, for poorly resolved galaxies, concentration index and $M_{20}$ suffer from strong discontinuities, i.e. similar morphologies are not necessarily mapped to neighbouring points in the parameter space. This limits the reliability of these parameters for classification purposes. Two-dimensional S\'ersic profiles accounting for centroid and ellipticity are identified as the currently most reliable parametrisation scheme in the regime of intermediate signal-to-noise ratios and resolutions, where asymmetries and substructures do not play an important role. We argue that basis functions provide good parametrisation schemes in the regimes of high signal-to-noise ratios and resolutions. Concerning S\'ersic profiles, we show that scale radii cannot be compared directly for profiles of different S\'ersic indices. Furthermore, we show that parameter spaces are typically highly nonlinear. This implies that significant caution is required when distance-based classificaton methods are used.
\end{abstract}


\section{Introduction}

In the last ten years the field of galaxy evolution has experienced a boost. With the advent of large ground-based spectroscopic and imaging surveys such as the SDSS \citep{Abazajian2009} or space-based surveys like COSMOS \citep{Scoville2007}, the database of galaxies has increased enormously. From both very deep as well as very wide area surveys substantial amounts of data are available, enabling us to study the dependence of galaxy formation and evolution on e.g. environment, star formation history or stellar/bulge/black hole mass. It is now possible to test multivariate dependencies and, in conjunction with numerical simulations, to describe possible evolutionary tracks of galaxies, and to single out not yet fully understood phenomena like the colour-bimodality of galaxies \citep[e.g.][]{Strateva2001} or the linear relation between black hole and stellar bulge mass \citep[e.g.][]{Haering2004,Woo2006}.

Studies of galaxy morphologies are very important in this context, because different morphologies are caused by different physical processes that are likely to also affect other properties, e.g. star-forming rate, and may also correlate with environment. Despite these efforts, it is still a very challenging task to meaningfully describe (parametrise) the morphologies of galaxies in very large data samples. Although we are well able to parametrise the morphologies of individual galaxies of certain types \citep[e.g.][]{Simmat2010}, finding a parametrisation scheme that is able to account for the huge variety of galaxy morphologies is a completely different task.

\subsection{Strategy}

In this paper we discuss the concept of parametrisation and summarise commonly used parametrisation schemes, namely CAS \citep{Abraham1994,Abraham1996,Bershady2000}, $M_{20}$ \citep{Lotz2004}, Gini \citep{Lotz2004,Lotz2008}, S\'ersic profile \citep{Sersic1968,Graham2005}, shapelets \citep{Refregier2003} and s\'ersiclets \citep{Ngan2009}. We categorise these schemes and identify important differences. However, the main intention of this article is to determine if there are any fundamental problems involved in the parametrisation of galaxy morphologies, which may turn out to be subtle or non-obvious. Our investigations are designed to test the current paradigm favouring model-independent schemes. It has already been shown that the diagnostic power of shapelets is limited for elliptical galaxies \citep{Melchior2009a}, whereas the method of s\'ersiclets has not yet been successfully established. Therefore, we focus our attention on the caveats involved in the usage of the other parametrisation schemes. In the course of this investigation, we demonstrate that morphological observables are intertwined. This new insight implies that all schemes that try to estimate observables separately without addressing their inherent degeneracies are problematic in principle.

In the remaining part of this introduction, we define the terms ``galaxy morphology'' and ``parametrisation'' and discuss what parametrisation is meant to achieve. In Sect. 2 we introduce two conceptually different approaches to parametrisation, namely model-independent (CAS, $M_{20}$, Gini) and model-based schemes (S\'ersic profile, shapelets, s\'ersiclets). As a first fundamental problem and one of our main results, we illustrate in Sect. 3 that morphological observables are intertwined and cannot be measured independently. Second, we investigate the impact of the point-spread function on the concentration index in Sect. 4. Third, we consider general problems affecting the classification of galaxy morphologies in Sect. 5. Finally, in Sect. 6 we summarise our results and give recommendations for improvements of existing or the design of new parametrisation schemes.

\subsection{Galaxy morphology\label{sect:galaxy_morphology}}

The morphology of a galaxy is defined by the characteristics of its two-dimensional light distribution, i.e. by the projected shape of the galaxy. Some morphological observables are:
\begin{itemize}
\item steepness of radial light profile
\item ellipticity (i.e. orientation \& axis ratio)
\item asymmetry (e.g. lopsidedness)
\item substructures (e.g. spiral arm patterns, bars, etc.)
\item size
\item centroid
\end{itemize}
The centroid position is an important morphological observable as well, since it is often required to derive other morphological estimators (cf. Table \ref{tab:para_schemes}). For decades galaxy morphologies have been studied in the visual regime, where all these observables are reasonably well defined. However, with increasing observational coverage of the electromagnetic spectrum, it became evident that morphology is a strongly varying function of wavelength. For instance, in the UV we observe mostly star-forming regions but no dust emission, such that galaxies can look patchy and highly irregular. On the other hand, in the far infra-red, there is almost no stellar but only dust emission. As we discuss in Sect. \ref{sect:assumptions}, many parametrisation schemes for galaxy morphologies make rather restrictive assumptions that are too specialised on the visual regime and cannot be generalised to the whole electromagnetic spectrum. As our discussion is set in the context of large surveys where galaxies exhibit a huge variety of different morphologies, we have to look for parametrisation schemes that are flexible enough to describe \textit{arbitrary} morphologies.

\subsection{Observation, parametrisation, inference\label{sect:trinity}}

\begin{figure}
\begin{center}
	\includegraphics[width=50mm]{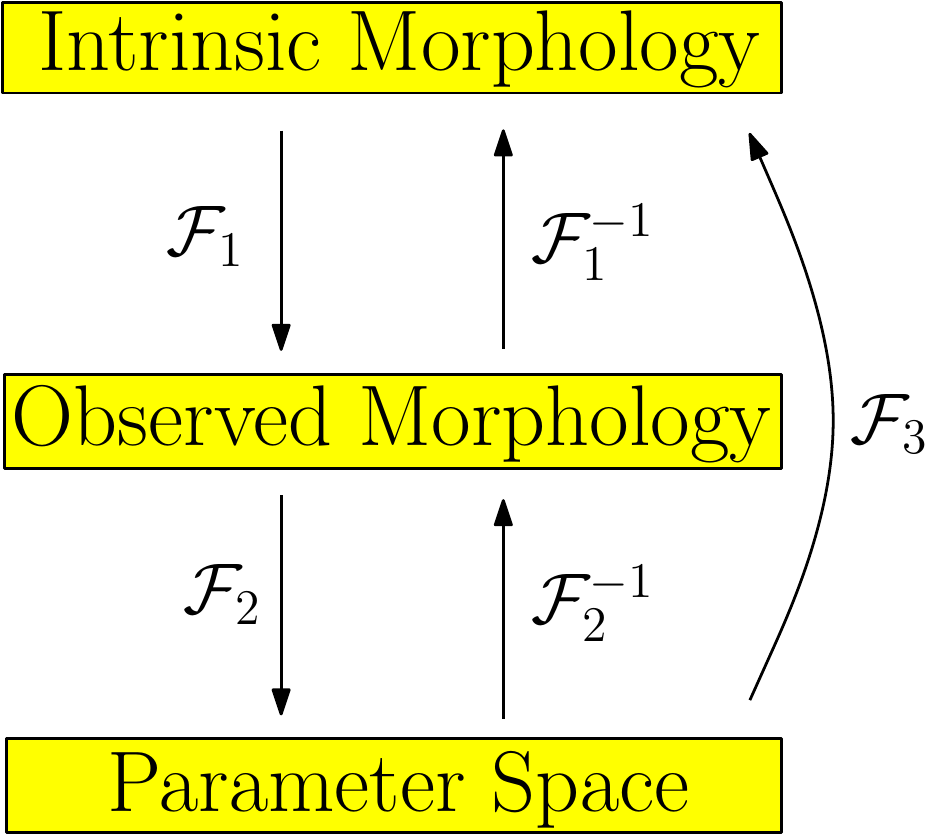}
\end{center}
\caption{Interplay of observation ($\mathcal F_1$), parametrisation ($\mathcal F_2$) and inference ($\mathcal F_3$). This paper is concerned with the existence of the mapping $\mathcal F_2^{-1}$, which is necessary for $\mathcal F_3=\mathcal F_1^{-1}\circ\mathcal F_2^{-1}$ to exist.}
\label{fig:trinity_observ_para_inf}
\end{figure}

In this section we want to clarify the role of parametrisation, i.e. what purpose it serves and what its benefits are. Parametrisation is one step in the sequence of observation, parametrisation and inference, which is visualised in Fig. \ref{fig:trinity_observ_para_inf}. 

The process of observation ($\mathcal F_1$) provides a nonlinear mapping of the true intrinsic galaxy morphology to an observed morphology. This mapping $\mathcal F_1$ comprises the projection onto the two-dimensional sky, the binning to pixels, the addition of pixel noise, and the convolution with the pixel-response function (gain of the detector). It also involves the convolution with the point-spread function, taking into account seeing effects, optics and instrument sensitivity.

However, analysing galaxy morphologies directly in pixel space is infeasible, since the number of pixels is typically very large. Therefore, it is necessary to parametrise the observed morphology ($\mathcal F_2$ in Fig. \ref{fig:trinity_observ_para_inf}), a step that has the two following aims: First, we want to reduce the degrees of freedom, since there is a lot of redundant or uninteresting information in pixel space. Second, we want to move from pixel space to some other description that better suits a given physical question. Effectively, this means that parametrisation can act as a method to reduce the dimensionality of the problem, to suppress noise and to extract information. Note that this definition of parametrisation encompasses more than just data modelling.

Based on such a parametrisation we can then try to infer the true intrinsic morphology. For instance, inference can be based on the search of multivariate dependencies of morphological descriptors on physical parameters or on classification. The inference step corresponds to the mapping $\mathcal F_3$ in Fig. \ref{fig:trinity_observ_para_inf}, where obviously $\mathcal F_3=\mathcal F_1^{-1}\circ\mathcal F_2^{-1}$, i.e. both mappings $\mathcal F_1$ and $\mathcal F_2$ need to be invertible -- at least in a practical sense. Often inference does not aim at the true intrinsic morphology, but at some abstract type or class that represents a reasonable generalisation. Still, if either $\mathcal F_1$ or $\mathcal F_2$ destroys too much information, this type of inference is impossible as well. For $\mathcal F_1$ being (approximately) invertible, the observation has to have a high signal-to-noise ratio and a high resolution relative to the features of interest (critical sampling). If this requirement is not met by the data, the observation will not resemble the true morphology and inference will be impossible. \citet{Bamford2009} observe this problem in the Galaxy Zoo project and term it ``classification bias''. They noticed that type fractions resulting from visual classifications of 557,681 SDSS galaxies with redshifts $z<0.25$ evolve significantly with $z$. As \citet{Bamford2009} do not expect a pronounced morphological evolution in this redshift regime, they assign this effect to the degradation of image quality with increasing redshift.

If $\mathcal F_2$ is invertible -- i.e. whether or not $\mathcal F_2^{-1}$ and thus $\mathcal F_3=\mathcal F_1^{-1}\circ\mathcal F_2^{-1}$ exists -- depends on the parametrisation scheme. This is the topic of this paper. Consequently, a reliable parametrisation is as important for inference as sufficient data quality.

\section{Parametrisation schemes\label{sect:para_schemes}}

In order to assess the advantages and deficites of different parametrisation schemes we now briefly summarise the most common approaches. We divide them into model-independent and model-based approaches. The most important difference is that the model-based approaches try to \textit{model} the two-dimensional light distribution of an image and are thus mostly descriptive. Model-independent approaches more directly try to extract physical information, hence mixing decription and inference steps. We conclude this section by summarising the assumptions involved in the parametrisation schemes.

\subsection{Model-independent schemes}

The foremost reason to use a model-independent -- or ``non-parametric'' -- approach is that it appears to be very simple at first glance. Most of these parametrisation schemes seem easy to implement, since they do not require to fit a model. Furthermore, parameters in all these schemes have at least a rough physical interpretation.


\subsubsection{CAS system}

A widely used set of morphological parameters is provided by the CAS system, which is based on the so-called Concentration, Asymmetry and Clumpiness indices  \citep{Abraham1994,Abraham1996,Bershady2000}. The concentration index is defined as
\begin{equation}\label{eq:def:concentration}
C = 5\,\log_{10}\left( \frac{r_{80}}{r_{20}} \right) \;\textrm{,}
\end{equation}
where $r_{80}$ and $r_{20}$ are the radii of circular (or elliptical) apertures containing
80\% and 20\% of the total image flux.\footnote{There are several variations of the concentration index: Sometimes it is based on the ratio of $r_{90}$ and $r_{50}$. Some authors \citep[e.g.][]{Bershady2000} consider the whole image for estimating $C$, others \citep[e.g.][]{Scarlata2007} estimate $C$ within a region given by one Petrosian radius.} The asymmetry index is defined as
\begin{equation}\label{eq:def:asymmetry}
A = \frac{\sum_\textrm{pixels}|I(\vec x) - I^{\textrm{180}^\circ}(\vec x)|}{\sum_\textrm{pixels}I(\vec x)} \;\textrm{,}
\end{equation}
where $I^{\textrm{180}^\circ}$ denotes the image $I$ rotated by
$\textrm{180}^\circ$. Obviously, the asymmetry $A$ is bound in the interval $[0,2]$. Finally, the clumpiness is defined as
\begin{equation}\label{eq:def:clumpiness}
S = 10\frac{\sum_\textrm{pixels}|I(\vec x) - I^\sigma(\vec x)|}{\sum_\textrm{pixels}I(\vec x)} \;\textrm{,}
\end{equation}
where $I^\sigma$ has been convolved by a Gaussian of width $\sigma$. The specific choice of $\sigma$ is somewhat arbitrary within a certain range, being sensitive to clumps of varying spatial extent. As far as we know, there is no systematic investigation of the impact of the choice of $\sigma$ on the parametrisation results.


\subsubsection{$M_{20}$ and Gini}

Two further morphological parameters are $M_{20}$ and the Gini coefficient. We define the second-order moment of pixel $n$ with value $I_n$ at position $\vec x_n$ as \citep{Lotz2004}
\begin{equation}\label{eq:def:2nd_moment_of_pixel}
M_n = I_n\,\left(\vec x_n - \vec x_c\right)^2 \;\textrm{,}
\end{equation}
where $\vec x_c$ denotes the reference position. Summation of the $M_n$ over all pixels yields the total second moment $M_\textrm{tot}$ with respect to $\vec x_c$. There is a theoretical preference to choose the reference position $\vec x_c$ to be the centre of light, because this choice minimises $M_\textrm{tot}$. $M_{20}$ is defined as
\begin{equation}\label{eq:def:M20}
M_{20} = \log_{10}\left(\frac{\sum_i M_i}{M_\textrm{tot}}\right) \;\textrm{,}
\end{equation}
where the summation $\sum_i M_i$ is over the pixels in descending order
$I_1\geq I_2\geq\ldots\geq I_N$ and stops as soon as $\sum_i I_i\geq
0.2\sum_{n=1}^N I_n$, i.e.\ as soon as 20\% of the total flux is reached. $M_{20}$ is supposed to estimate the spatial distribution of the most luminous parts of a galaxy image.

The Gini coefficient was defined by \citet{Lotz2004,Lotz2008} based on \citet{Glasser1962} as
\begin{equation}\label{eq:def:gini}
G = \frac{\sum_{n=1}^N (2n-N-1)|I_n|}{(N-1)\sum_{n=1}^N |I_n|} \;\textrm{,}
\end{equation}
where $N$ is the number of image pixels and $|I_1|\leq |I_2|\leq\ldots\leq |I_N|$ are the absolute values of the pixel fluxes sorted in ascending order. In contrast to $M_{20}$, Gini does not require an estimate of the centroid position. The Gini coefficient estimates the distribution of the pixel values over the image. As shown by \citet{Lisker2008}, it strongly depends on the signal-to-noise distribution within a galaxy's image and is thus a highly unstable morphological estimator.


\subsection{Model-based schemes I: S\'ersic profile\label{sect:sersic-index}}

\subsubsection{Definition}

The radial light profiles of many galaxies are reasonably well described by the S\'ersic profile \citep[see][for a compilation of relevant formulae]{Sersic1968,Graham2005},
\begin{equation}\label{eq:def:Sersic_model}
I(R) = I_\beta\,\exp\left\{-b_n\left[ \left(\frac{R}{\beta}\right)^{1/{n_S}} -1 \right]\right\} \;\textrm{,}
\end{equation}
where $n_S$ is the S\'ersic index and $\beta$ is the scale radius\footnote{The scale radius $\beta$ is expressed in units of pixels, i.e. $\beta^{-1}$ is the pixel size relative to the object size.}. The constant $b_n$ is usually chosen such that the radius $\beta$ encloses half of the total light. $I_\beta$ is the intensity at the half-light radius $\beta$. At fixed $n_S$, $b_n$ is then given by
\begin{equation}\label{eq:definition_b_n}
\Gamma(2n_S) = 2\gamma(2n_S,b_n) \;\textrm{,}
\end{equation}
where $\Gamma$ and $\gamma$ denote the complete and incomplete gamma functions. For $n_S>0.5$ one can approximate $b_n\approx 2 n_S - \frac{1}{3}$. The S\'ersic profile corresponds to a Gaussian profile if $n_S=0.5$, to an exponential disk profile if $n_S=1$, and to a deVaucouleur profile if $n_S=4$.

Throughout this paper we use a truncated S\'ersic profile of the form
\begin{equation}
\tilde I(R) = \left\{\begin{array}{lcr}
I(R) - I(5\beta) & \Leftrightarrow & R \leq 5\beta \\
0 & \textrm{otherwise} &
\end{array}\right. \;\textrm{,}
\end{equation}
such that all profiles are 0 for $R>5\beta$ but still continuous. This is necessary, since otherwise the profiles do not vanish quickly enough for large S\'ersic indices.

\subsubsection{Redefining $b_n$ and $\beta$}

\begin{figure}
      \includegraphics[width=84mm]{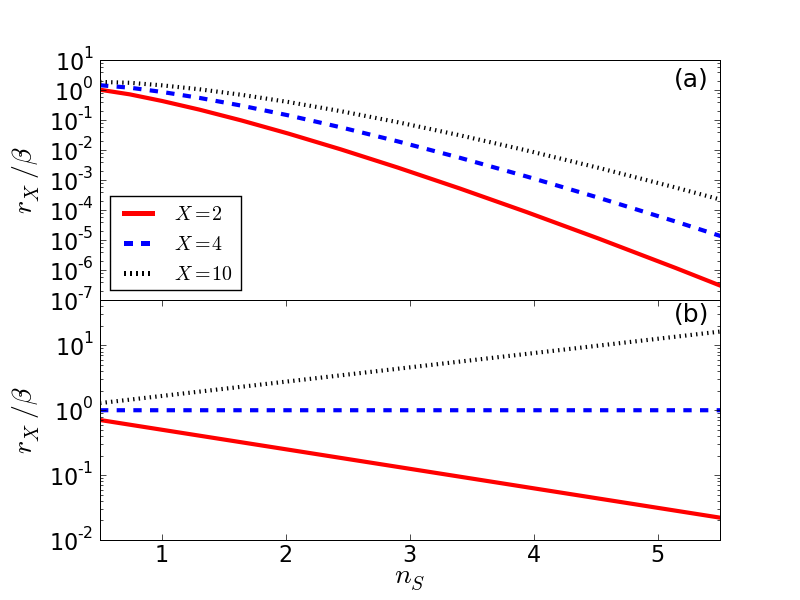}
\caption{Radii $r_X$ where S\'ersic profile takes values $I(r_X)/I(0)=1/X$ for $X=2,4,10$ and $b_n$ given by Eq. (\ref{eq:definition_b_n}) (panel (a)) and $b_n=\log 4$ (panel (b)).}
\label{fig:interpretation_scale_radius_4_n_S}
\end{figure}

\begin{figure}
      \includegraphics[width=84mm]{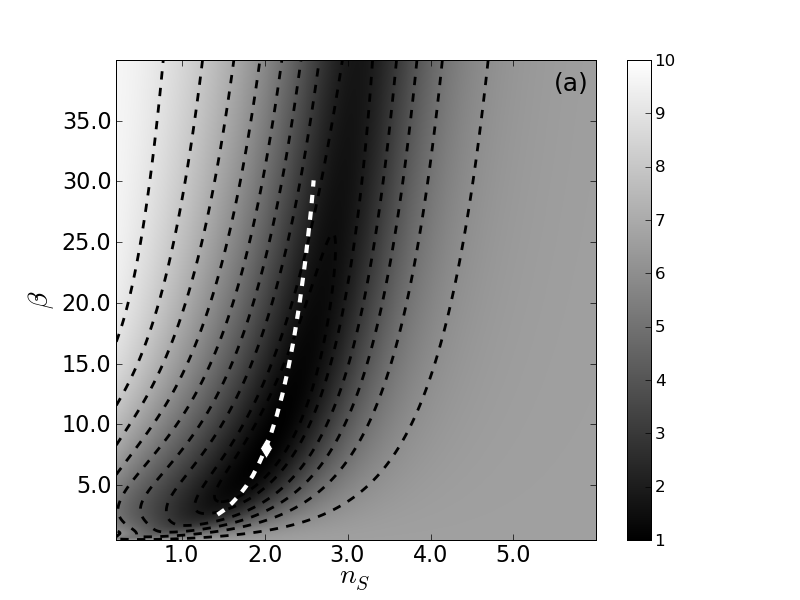}
      \includegraphics[width=84mm]{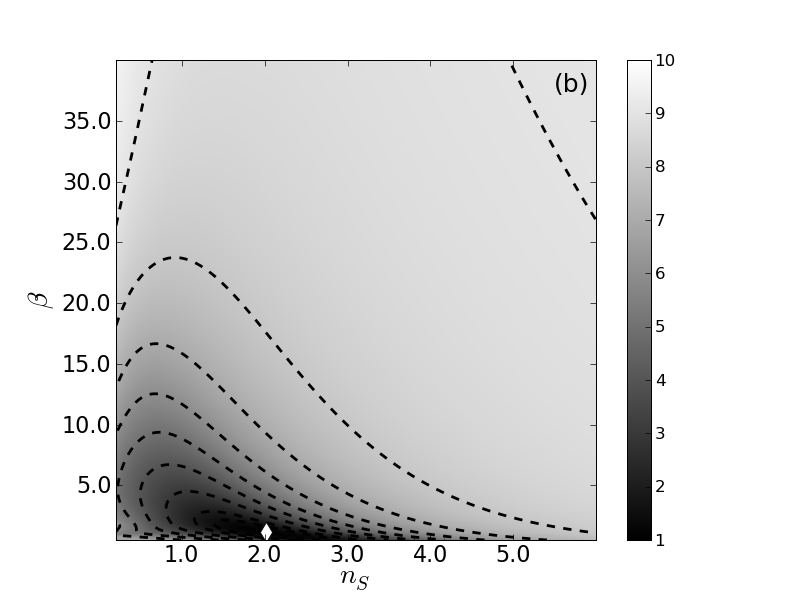}
\caption{$\chi^2/\textrm{dof}$ manifolds demonstrating how Eq. (\ref{eq:definition_b_n}) induces the artificial correlation of $n_S$ and $\beta$.\newline
(a) $\chi^2/\textrm{dof}$ manifold for $b_n$ defined by Eq. (\ref{eq:definition_b_n}). The white diamond indicates the optimum. The dashed white line is given by $b_n/\beta^{1/n_S}=\textrm{const}$ and follows the valley, thereby illustrating that the correlation of $n_S$ and $\beta$ is artificial.\newline
(b) Same as in (a) but for $b_n=\log 4$ for all $n_S$. The valley is approximately parallel to the $n_S$-axis, i.e. the correlation is gone.\newline
Both panels use the same artificial light profile with low noise level to evaluate the $\chi^2/\textrm{dof}$ manifold. It is much easier to find the optimum in panel (b) than in panel (a). The optimal values of $n_S$ are identical in (a) and (b), whereas the optimal values of $\beta$ are different due to the different choice of $b_n$. $\chi^2/\textrm{dof}$ is not a simple quadratic form, because the S\'ersic profile is a nonlinear model.}
\label{fig:b_n_inducing_artificial_correlation}
\end{figure}

It is important to note that $b_n$ and $\beta$ in Eq. (\ref{eq:def:Sersic_model}) are completely degenerate. We are free to make any choice of $b_n$ that is different from Eq. (\ref{eq:definition_b_n}), thereby redefining the model and changing the meaning of $\beta$. There are two reasons why Eq. (\ref{eq:definition_b_n}) potentially is not a good choice for $b_n$:
\begin{enumerate}
\item From a theoretical point of view, half-light radii $\beta$ are \textit{not} comparable for different values of $n_S$, i.e. the size of one galaxies relative to a second galaxy can be inferred from their scale radii if \textit{and only if} both S\'ersic models use identical $n_S$. However, in practice this is rarely a problem, since studies usually compare only sizes of galaxies of similar Hubble types, e.g.\ in studies of the size-evolution of disc galaxies. Nonetheless, choosing $b_n$ according to Eq. (\ref{eq:definition_b_n}), we \textit{must not} demand that $\beta$ is smaller than the image size, since $\beta$ \textit{cannot} be interpreted this way. We actually need to require that the profile drops within the image boundaries. Figure \ref{fig:interpretation_scale_radius_4_n_S} shows that the radii where the profile reduces to $\frac{1}{2}$, $\frac{1}{4}$, and $\frac{1}{10}$ of its value at $r=0$ vary over several orders of magnitude for different $n_S$. The scale radius $\beta$ is more intuitively defined such that
\begin{equation}\label{eq:def_1oX}
\frac{I(\beta)}{I(0)} = 1/X
\end{equation}
for some $X>0$ \textit{independent} of $n_S$. This can be achieved by setting $b_n=b=\log X$ for all $n_S$. Panel (b) of Fig. \ref{fig:interpretation_scale_radius_4_n_S} shows that in this case the radii for different $n_S$ change by less than two orders of magnitude and hence can be compared much better.
\item It is well known that there is a strong correlation of $n_S$ and $\beta$ \citep[e.g.][]{Trujillo2001}, which is problematic for many fit algorithms. This correlation of $n_S$ and $\beta$ is almost completely induced by Eq. (\ref{eq:definition_b_n}), i.e. it is artificial. We can remove this correlation by setting $b_n=\log X$ for all $n_S$, thereby simplifying the fit problem. We demonstrate this in Fig. \ref{fig:b_n_inducing_artificial_correlation} showing $\chi^2$ manifolds for fitting an artificial light profile once using Eq. (\ref{eq:definition_b_n}) (panel a) and once using $b_n=\log X$ for all $n_S$ (panel b). The noise level in this simulation is low (the signal-to-noise ratio of the central peak is 100). Higher noise levels will not change the curvatures of the $\chi^2$ ``valleys'' in Fig. \ref{fig:b_n_inducing_artificial_correlation} but will only broaden them and reduce their depth.
\end{enumerate}
These issues are not fundamental and there is no theoretical preference for choosing between these approaches apart from the fact that Eq. (\ref{eq:def_1oX}) is likely to provide more robust parameter estimates. Furthermore, it is possible to convert to and fro the definitions of Eqs. (\ref{eq:definition_b_n}) and (\ref{eq:def_1oX}) via $b/\beta^{1/n_S}=\textrm{const}$.

\subsection{Model-based schemes II: Expansion into basis functions\label{sect:basis_functions}}

An alternative model-based parametrisation approach is the expansion into basis functions. The most important advantage of this concept is that the parametrisation is more flexible, whereas all previous schemes are highly specialised for certain morphologies. A good set of basis functions should be able to fit almost anything, provided the signal-to-noise ratio of the given data is sufficiently high. Hence, this approach should in principle be favoured when the task at hand is to parametrise arbitrary morphologies.

Basis-function expansions are very common in physics and also in cosmology (e.g. decomposing the CMB into spherical harmonics). Usually, the basis functions are chosen based on symmetry arguments or best as eigenfunctions of the differential equations describing the underlying physics. However, we do not know the physics governing galaxy morphologies yet, hence there is no theoretically motivated choice for the set of basis functions. Therefore, basis functions are chosen such that they possess advantageous analytic properties or overcome special problems.

In the following, we introduce the concept of basis functions. We briefly comment on the issues of orthonormality and completeness and then discuss example sets of basis functions.


\subsubsection{General concept}\label{sect:concept_linear_basis_functions}

A set of basis functions is usually defined such that it is orthonormal and complete. However, we want to introduce this concept in a sligthly more general fashion. Consider a set of $N$ scalar-valued functions $\{B_1(\vec x;\vec\theta_1),\ldots,B_N(\vec x;\vec\theta_N)\}$, where $\vec x$ denotes the two-dimensional pixel-position vector and $\vec\theta_n$ is the set of parameters of the $n$-th basis function $B_n$. The basis functions may be nonlinear in both $\vec x$ and $\vec\theta_n$. We consider the linear superposition, i.e. the model,
\begin{equation}\label{eq:def:linear-model}
f(\vec x) = \sum_{n=1}^N c_n B_n(\vec x;\vec\theta_n) \;\textrm{,}
\end{equation}
with the $N$ expansion coefficients $c_n$. These coefficients are further model parameters in addition to $\vec\theta_n$. The $c_n$ enter Eq. (\ref{eq:def:linear-model}) linearly, hence they form a linear space, i.e.\ a vector space. Therefore, the set of $N$ coefficients is also referred to as ``coefficient vector'' $\vec c$. Given an observed galaxy image $I(\vec x)$, we can fit the model $f(\vec x)$ to this image. The details of the fitting process will depend on the choice of the set of basis functions. The fitting process itself is also called the ``decomposition of the image into the basis functions''.

After fitting the model defined by Eq. (\ref{eq:def:linear-model}) to the image, we obtain estimates for the coefficients $c_n$ and the parameters $\vec\theta_n$ for all basis functions. Usually, the $\vec\theta_n$ are used to incorporate several effects. For instance, there is typically a size parameter that scales the spatial extent of the basis functions such that the coefficients $c_n$ do not depend on the size of the object. If this is the case, then the basis functions are called ``scale invariant''. The centroid position can also be part of $\vec\theta_n$. The linear coefficients $c_n$ are supposed to capture the morphological information.


\subsubsection{Orthonormality and completeness\label{sect:orthogonality-completeness}}

As aforementioned, sets of basis functions are often orthonormal and complete. The orthogonality would ensure that all coefficients were completely independent of each other. The completeness would allow us to decompose an {\it arbitrary} image. In practice, however, the completeness is lost due to pixel noise and pixellation, which sets an upper limit to the number of basis functions that can be used to decompose a given image. This can lead to characteristic modelling failures. We discuss this in slightly more detail in the next section. The strict orthogonality is also lost, due to pixellation \citep{Melchior2007}. This means that the resulting coefficients may exhibit minor correlations, but if the galaxy image and all basis functions are critically sampled, these correlations will be negligible.

\subsubsection{Shapelets}

Shapelets were introduced by \citet{Refregier2003}. They are a scaled version of Gauss-Hermite polynomials, i.e.
\begin{equation}\label{eq:def:shapelets}
B_n(x; \beta) = \left( 2^n n! \sqrt{\pi}\beta \right)^{-1/2} H_n\left(\frac{x}{\beta}\right)\exp\left[-\frac{x^2}{2\beta^2}\right] \;\textrm{,}
\end{equation}
where $H_n$ denotes the Hermite polynomial of order $n$ and $\beta$ is the shapelet scale size. A centroid can be introduced via $x\rightarrow x-x_0$. In this case, all basis functions take identical parameters $\vec\theta_n=\vec\theta=(x_0,\beta)$ in order to allow for orthogonality. From this definition, we can build two-dimensional basis functions, namely Cartesian shapelets and polar shapelets.

The Gaussian weight function of shapelets leads to very nice analytical properties. For instance, shapelets are nearly invariant under Fourier transformation, which makes any convolution or deconvolution a closed and analytic operation in shapelet space, as described in \citet{Melchior2009}. However, the limitation of basis functions due to pixel noise has a severe consequence: Shapelets employ a Gaussian weight function (cf. Eq. (\ref{eq:def:shapelets})), but real galaxies have typically much steeper profiles. This gives rise to characteristic modelling failures that typically manifest themselves in ring-like artifacts in the shapelet reconstructions of galaxies with exponential or steeper light profiles. This severly limits the diagnostic power of shapelets \citep[cf.][]{Melchior2009a} and we therefore exclude them from our subsequent simulations.

Despite these fundamental problems, shapelets demonstrate a very important aspect of basis-function expansions: For highly resolved galaxies of high signal-to-noise ratios S\'ersic profiles are incapable of providing excellent models as they are not flexible enough to account for substructures such as spiral arm patterns, i.e. their residuals do not always reach noise level. In case of shapelets -- as an example of basis functions -- this is fundamentally different. They are highly flexible and reach noise level even for galaxies that are very large, highly resolved and bright \citep[e.g.][]{Andrae2010a}.

\subsubsection{S\'ersiclets}

Given the problematic impact of the Gaussian profile on shapelets, a set of basis functions based on the S\'ersic profile is an obvious means to overcome the limitations of shapelets. The resulting basis functions are called s\'ersiclets. \citet{Ngan2009} were the first to realise the potential of this approach, which is capable of accounting for all morphological observables listed in Sect. \ref{sect:galaxy_morphology}. However, for technical reasons their implementation of s\'ersiclets was flawed, as we illustrate in an upcoming paper \citep{Andrae2010c}. We therefore also exclude s\'ersiclets from our simulations.

\subsubsection{Outlook: Template libraries}

We already argued that no basis set -- apart from the pixel grid itself -- is actually complete due to the limitations induced by pixel noise. Now, we want to briefly touch -- without going into details -- on a set of basis functions that is finite and thus incomplete from the beginning. The motivation is very simple: For both shapelets and s\'ersiclets the basis functions lack a physical interpretation. Why not use basis functions that directly correspond to spiral arms, galactic bars or rings? We can use a set of such \textit{templates} -- a template library -- to form linear models and decompose the image, resulting in a set of coefficients that form a vector space. The individual templates do not even need to be orthogonal, but just as linearly independent as possible in order to avoid heavy degeneracies during the fitting procedure. Unfortunately, the direct physical motivation is also the major drawback of this approach, since we are strongly prejudiced and lack flexibility in this case. For instance, template libraries are likely to have severe problems in decomposing irregular galaxies, i.e. they are inappropriate for parametrising arbitrary morphologies. Moreover, the set of morphological features is very large, hence such a library has to contain numerous templates.

\subsection{Assumptions\label{sect:assumptions}}

It is crucial to be aware of all assumptions made by a certain method when using it, since if a method fails, it usually fails because one or more of its assumptions break down. In case of model-based approaches, the assumptions are usually rather obvious and therefore can be easily challenged. In contrast to this, the assumptions of model-independent approaches are implicit and often hidden. This may lead to the misapprehension that model-independent schemes were superior since they required fewer or even no assumptions.

In Table \ref{tab:para_schemes} we summarise our categorisation of parametrisation schemes. Based on this table and the definitions given in the previous sections, we now work out the assumptions of all schemes from a \textit{theoretical} point of view. In practice, it is virtually impossible to satisfy all assumptions. Whether the violation of some assumption leads to a breakdown of a certain method depends on the specific question under consideration, the desired precision, the details of the method's implementation, and the quality of the data. In detail, the assumptions are:
\begin{itemize}
\item Concentration index: There are no azimuthal structures such as spiral-arm patterns or galactic bars.\footnote{
This is a mathematical and deeply implicite assumption that is generally not realised when working with actual galaxy data: The ``radii'' used to compute the concentration index are estimated from a curve of growth. This curve of growth is actually a two-dimensional \textit{integral} over the galaxy's light profile (though it is usually reduced to a summation due to pixellation to allow a comment on an actually irrelevant practical detail). Nevertheless, it is inevitable to parametrise this integration in some way in order to be \textit{capable} of evaluating it (analytically or numerically). In simple words, one has to \textit{define} what ``radius'' means (e.g. spherical or elliptical radius) and this definition is the assumption. For instance, assuming spherical integration contours, the curve-of-growth integral of an image $f(r,\varphi)$ reads $p(R)=\int_0^R dr\,r\int_0^{2\pi}d\varphi\,f(r,\varphi)$, where the integral has been parametrised in polar coordinates. In fact, Figure \ref{fig:impact_eps_on_C} can be understood as investigating what happens if the curve of growth indeed takes this spherical form but the image data is not spherically symmetric but elliptical. More physically, though already beyond the point: In case of an image that has perfectly circular or elliptical symmetry, the azimuthal integration in $p(R)$ is well defined and so are the radii and the concentration index. However, if there is more complicated azimuthal structure than ellipticity, there is no simple way to generally define the curve of growth. Either, the integration is along true isophotes. In this case, the shape of the integration regions will vary from object to object and potentially also with radius. Then the resulting concentration indices would not be comparable. The other option is to integrate along given circular or elliptical isophotes to enforce comparability. This approach explicitely assumes that there is no azimuthal structure or else the radius in $p(R)$ has no strict relation to the galaxy, and the estimated curve of growth will be biased. The justification to use this in practice is to assume that in reality objects of similar type will catch similar biases, such that concentration indeces still have discriminative power in a differential sense, though their absolute values may be biased. Furthermore, the mere presence of such a bias does not automatically imply that the resulting estimates of the curve of growth and the concentration index, respectively, are not meaningful anymore.
} The pixel noise is negligible and the object is not grossly asymmetric such that a centroid is well defined (cf. Sect. \ref{sect:asymmetry_vs_centroid}). The scheme can be enhanced using elliptical apertures.
\item Asymmetry index: A centre of rotation is well defined. The pixel noise is negligible. Both issues have been addressed by \citet{Conselice2000b}. The asymmetry of interest is visible under rotations of $180^\circ$.
\item Clumpiness index: The functional type of the kernel matches the galaxy profile. The width of the kernel is chosen such that the information of interest is extracted. The ellipticity of the kernel matches the ellipticity of the object.
\item $M_{20}$: The pixel noise has negligible impact on the estimates of centroid and second moments. The centre of light and the object's centre coincide, i.e. there is no substantial asymmetry. The structures dominating $M_{20}$ are of circular shape with the centroid at their centres.\footnote{This assumption stems from the term $(\vec x_n-\vec x_0)^2$ in Eq. (\ref{eq:def:2nd_moment_of_pixel}).}
\item Gini coefficient: The pixel noise is negligible \citep[see][]{Lisker2008}.
\item S\'ersic profile: The S\'ersic profile is a good match of the object's light profile. In particular, this means that the object's light profile is symmetric, monotonically decreasing and the steepness is correctly described by the model, and there are no azimuthal structures such as spiral arm patterns, galactic bars or rings.
\item (Spherical) shapelets: Employing the Gaussian weight function fits galaxy profiles. Using spherical basis functions that have no intrinsic ellipticity does not lead to problems.
\end{itemize}
We now clearly see that model-independent schemes implicitely make assumptions, too. This list suggests that non-parametric approaches \textit{tend} to invoke fewer assumptions than model-based schemes\footnote{However, it is \textit{not} true in general that model-independent schemes invoke fewer assumptions than model-based approaches. As an exception to this ``rule'', compare concentration index and shapelets.} at the loss of reliability, as we are going to demonstrate in the following sections. We also want to emphasise that shapelets -- as an example of basis functions -- can describe asymmetries.

\begin{table*}
\begin{tabular}{l|cccccccc}
\hline\hline
Characteristic & $C$ & $A$ & $S$ & $M_{20}$ & $G$ & S\'ersic profile & shapelets & s\'ersiclets \\
\hline
model-based & n & n & n & n & n & y & y & y \\
centroid estimate necessary & y & y & n & y & n & y & y & y \\
account for steepness of light profile & n & n & n & n & n & y & n & y \\
account for ellipticity & y${}^{(1)}$ & y${}^{(2)}$ & y${}^{(3)}$ & n & n & y & y/n${}^{(4)}$ & y \\
account for substructures & n & y & y & n & n & n & y & y \\
\end{tabular}
\caption{Characteristics of parametrisation schemes.\newline
${}^{(1)}$ We can employ elliptical isophotes to compute $C$.\newline
${}^{(2)}$ $A$ is invariant under all operations that are symmetric under rotations by $180^\circ$. Ellipticity is such an operation.\newline
${}^{(3)}$ It is possible to use an elliptical Gaussian for convolution.\newline
${}^{(4)}$ There are spherical and elliptical shapelet formalisms.}
\label{tab:para_schemes}
\end{table*}


\section{Intertwinement of morphological observables\label{sect:entanglement}}

The basic idea of model-independent schemes is to estimate the different morphological observables listed in Sect. \ref{sect:galaxy_morphology} independently of each other, thereby simplifying the problem. However, in this section we present as one of our main results the fact that these morphological observables are intertwined, which means that it is impossible to measure them independently of each other. Even if we try to measure only a single observable using a method unaware of the other observables, the mere presence of these observable features will influence the results. The notion of intertwinement should not be confused with redundancy, e.g. S\'ersic index and concentration index are perfectly redundant (Sect. \ref{sect:ZEST_bias}) but asymmetry and concentration index are not (Sect. \ref{sect:asymmetry_vs_centroid}). Of course, for some observables the intertwinement is stronger than for others. This intertwinement is not of physical origin but stems from the fact that usually all morphological observables are present simultaneously, such that the assumptions listed in Sect. \ref{sect:assumptions} are \textit{never} truly satisfied.

We carry out noise-free simulations of the different parametrisation schemes and by doing so we reveal several systematic misestimations -- in particular of the concentration index. All simulations invoke Sersic profiles and we want to explicitly emphasise that it is \textit{not} necessary for real galaxies to actually follow Sersic profiles.\footnote{To be more precise, it is perfectly valid to use such idealised simulations to discover these biases, but in order to correct for them more realistic simulations are necessary.} However, as we demonstrate in Sect. \ref{sect:ZEST_bias}, S\'ersic profiles provide parametrisations that are in excellent agreement with estimates of light concentration. This would not be the case if S\'ersic profiles were a bad description. Pixel noise in real data may hide these biases to some extent, but they will still be present.

\subsection{Example I: S\'ersic profile vs. concentration index\label{sect:ZEST_bias}}

\begin{figure}
      \includegraphics[width=84mm]{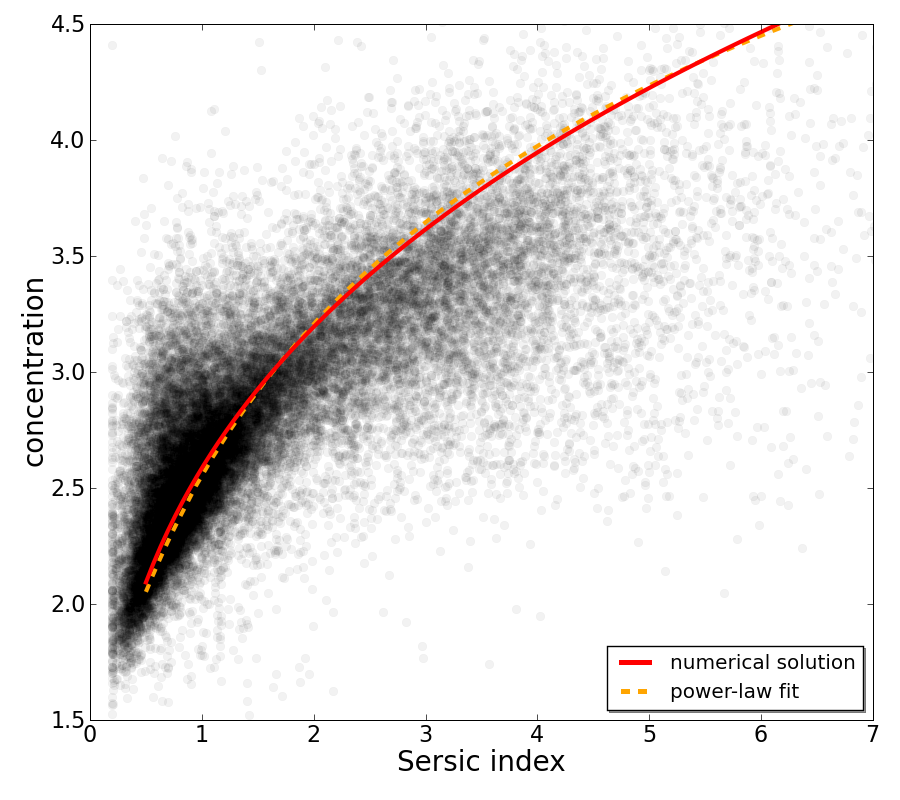}
\caption{Comparing concentration and S\'ersic indices of 31,288 COSMOS galaxies from the Zurich Structure \& Morphology catalogue \citep{Sargent2007} (blue points) with the numerical solution (red solid curve) and power-law fit of Eq. (\ref{eq:C_as_f_of_n_S_approx}) (oranged dashed curve). Shown are COSMOS galaxies with $I<22.5$, valid axis ratios ($0<q\leq 1$), and flags ``stellarity'', ``junkflag'' and ``flagpetro'' of 0. Concentration indices were predicted from analytic S\'ersic profiles using numerical integration out to one Petrosian radius. There was \textit{no} pixellation.}
\label{fig:C_vs_n_S_for_ZEST}
\end{figure}

We begin with comparing S\'ersic profiles and the concentration index, establishing a relation between both schemes that allows us to assess systematic effects on the concentration. The S\'ersic index estimates how steeply the radial light profile falls off. Consequently, S\'ersic index and concentration index are essentially two estimators for the same morphological feature, namely the steepness of the light profile. This is also evident from the fact that both schemes have almost identical assumptions (cf. Sect. \ref{sect:assumptions}). In fact, we can compute the concentration of a two-dimensional S\'ersic profile using numerical integration, i.e., S\'ersic index and concentration index are perfectly redundant \citep[see also][]{Trujillo2001}. Integration the flux to infinite radius, Eq. (\ref{eq:def:concentration}) yields the power law
\begin{equation}
C \approx 2.770\cdot n_S^{0.466} \;\textrm{,}
\end{equation}
which provides a good approximation for the exact numerical solution for $0.5\leq n_S\leq 7$. The resulting values of $n_S=0.5,1$ and 4 are identical to those given by \citet{Bershady2000}. Integration the flux to one Petrosian radius instead of infinity, the approximate solution is
\begin{equation}\label{eq:C_as_f_of_n_S_approx}
C \approx 2.586\cdot n_S^{0.305} \;\textrm{.}
\end{equation}
Obviously, any declining radial profile can be mapped onto the concentration index this way, irrespective of whether or not it is a good description of a galaxy. Therefore, Fig. \ref{fig:C_vs_n_S_for_ZEST} also compares this theoretical expectation with the measured concentration indices and S\'ersic indices of 31,288 COSMOS galaxies from the Zurich Structure \& Morphology catalogue \citep{Scarlata2007,Sargent2007}.\footnote{http://irsa.ipac.caltech.edu/data/COSMOS/datasets.html} Evidently, the \textit{independent} estimates of concentration indices conducted by \citet{Scarlata2007} and of S\'ersic indices conducted by \citet{Sargent2007} are in excellent agreement with the theoretical prediction of Eq. (\ref{eq:C_as_f_of_n_S_approx}). This clearly demonstrates that concentration and S\'ersic indices are equivalent parametrisations in case of COSMOS galaxies, providing largely unbiased estimates. Nevertheless, this single example does \textit{not} supersede a detailed study of potential biases that may occur in practice. In particular, the COSMOS data shown in Fig. \ref{fig:C_vs_n_S_for_ZEST} exhibits a large scatter that may hide biases.

\subsection{Example II: Steepness of light profile vs. ellipticity}

Our second example is the intertwinement of the steepness of the radial light profile and the ellipticity. These two are certainly the most important morphological observables listed in Sect. \ref{sect:galaxy_morphology}, having the largest impact on parametrisation results.

\begin{figure}
      \includegraphics[width=84mm]{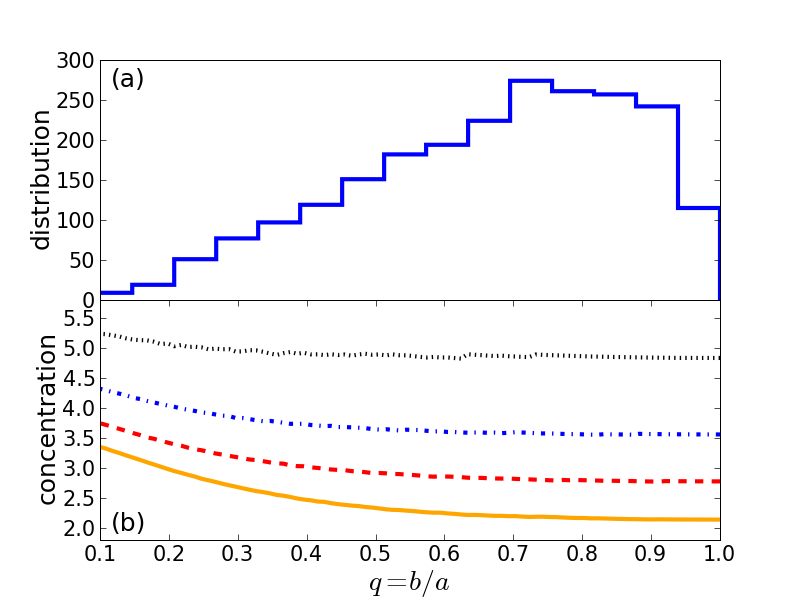}
\caption{Impact of ellipticity on concentration estimates. Panel (a) shows the distribution of axis ratios $q=b/a$ for 2,272 SDSS galaxies from the data sample of \citet{Fukugita2007}. Panel (b) shows concentration estimates using circular isophotes for elliptical S\'ersic profiles with $n_S=0.5$ (solid orange line), $n_S=1$ (dashed red line), $n_S=2$ (dotted-dashed blue line), and $n_S=4$ (dotted black line).}
\label{fig:impact_eps_on_C}
\end{figure}

It is obvious that estimates of the steepness of the radial light profile must take into account ellipticity. Therefore, it is necessary to use elliptical isophotes in case of the concentration index or to fit a two-dimensional S\'ersic profile that is enhanced by an ellipticity parameter. Unfortunately, in case of the SDSS, the aperture radii containing 50\% and 90\% of the total image flux given in the SDSS database are chosen as circular apertures \citep{Strauss2002}. This implies that estimates of the concentration index drawn from these values may be biased. In fact, this bias was already discussed by \citet{Bershady2000}. They investigated how the concentration index changes with axis ratio for samples of real galaxies of similar morphological types. \citet{Bershady2000} claim that using circular apertures causes an overestimation of concentration indices of at most 3\% and is therefore negligible. We investigate this effect in Fig. \ref{fig:impact_eps_on_C} for a realistic range of axis ratios, as is evident from panel (a). Panel (b) shows how the concentration index is influenced by the axis ratio for S\'ersic profiles with fixed S\'ersic indices, corresponding to galaxy samples of similar morphologies as in \citet{Bershady2000}.\footnote{Obviously, S\'ersic profiles are rather idealised and by far not as realistic as the sample used by \citet{Bershady2000}. However, this does \textit{not} hamper the validity of this test, but rather serves the purpose of isolating this bias. Apart from that, there is no difference in both studies.} Evidently, for $q\gtrsim 0.5$ -- which is the majority of galaxies in the given set -- the bias is negligible. There are galaxies with $q<0.5$, which are typically disc-like galaxies with shallow light profiles. For those objects concentration estimates based on circular isophotes are substantially overestimated ($\approx 30\%$ for $n_S=1$). This bias is \textit{not} negligible. \citet{Bershady2000} based their investigation on estimated concentration indices of \textit{real} galaxies. Hence, the most likely origin of this discrepancy in our results is that the intrinsic scatter in the real data used by \citet{Bershady2000} hid this bias. Considering ellipticity and concentration index together -- instead of using an elliptical concentration index -- is \textit{not} likely to solve this problem. The reason is that incorporating an ellipticity estimate may add information about the cause of the bias of the concentration index, but it does not provide information about the effect of this bias. Finally, we want to emphasise that Fig. \ref{fig:impact_eps_on_C} must not be used to calibrate the biased concentration estimates resulting from circular apertures. The reason is that this would now require S\'ersic profiles to be a realistic description of galaxy morphologies. Moreover, also the study of \citet{Bershady2000} cannot be used for such a purpose, because the bias clearly depends on the intrinsic concentration. This means that such a correction would require prior knowledge about the object's true concentration.

Vice versa, \citet{Melchior2009a} showed in the context of weak gravitational lensing that ellipticity measurements using shapelets are strongly biased in case of steep profiles. In other words, shapelets fail to provide reliable ellipticity estimates, because they do not properly account for the steepness of the radial light profile. This impressively demonstrates that these two observables may be closely intertwined.

\subsection{Example III: Impact of lopsidedness on centroid estimation\label{sect:asymmetry_vs_centroid}}

\begin{figure}
\begin{center}
      \includegraphics[width=35mm]{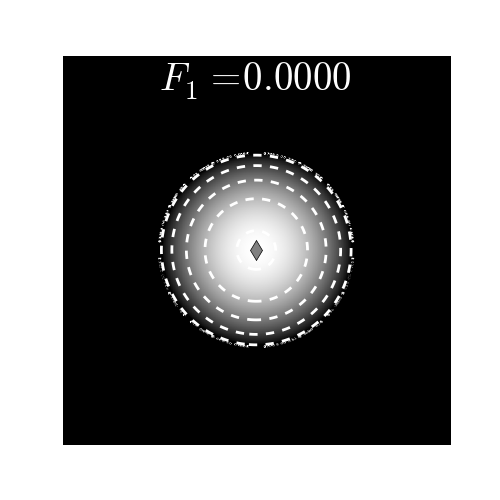}
      \includegraphics[width=35mm]{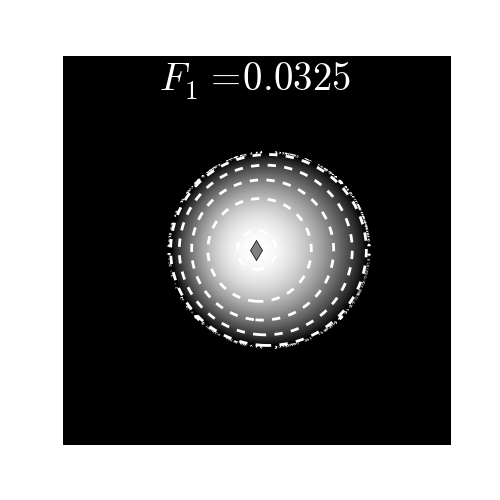}
      \includegraphics[width=35mm]{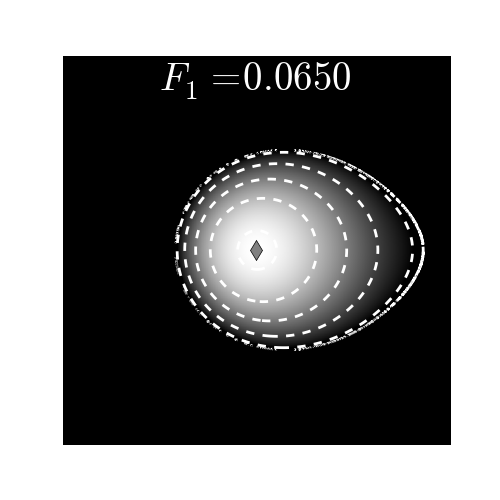}
\end{center}
\caption{Gaussian profiles of different lopsidedness. The applied flexions are $F_1=0.0$ (top left), $F_1=0.0325$ (top right), and $F_1=0.065$ (bottom). The resulting profiles exhibit realistic lopsidedness. All profiles are evaluated on a 1000$\times$1000 pixel grid using a scale radius of $\beta=50$. White diamonds indicate the maximum position.}
\label{fig:flexed_gaussians}
\end{figure}

As a third example for the intertwinement of morphological observables, we consider the impact of asymmetry on centroid estimates and the resulting parameter estimation using two-dimensional S\'ersic profiles. We simulate a certain type of asymmetry, namely lopsidedness. In order to introduce lopsidedness analytically, we apply the flexion transformation from gravitational weak lensing \citep{Goldberg2005} to the S\'ersic profiles as explained in Appendix \ref{app:shear_flexion_trafo}. The strength of the flexion transformation is parametrised by $F_1$, $F_2$, $G_1$, and $G_2$. There is no pixel noise in this simulation. Figure \ref{fig:flexed_gaussians} shows Gaussian profiles resulting from this transformation.\footnote{The flexion transformation of Eq. (\ref{eq:flexion_trafo}) will produce a second solution of $\vec x^\prime=0$, which corresponds to multiple images in weak lensing. We only consider cutouts with just one image, where the other image resulting from the second solution to $\vec x^\prime=0$ is far away.} The resulting distortions are not unrealistically strong.

\begin{figure}
\includegraphics[width=84mm]{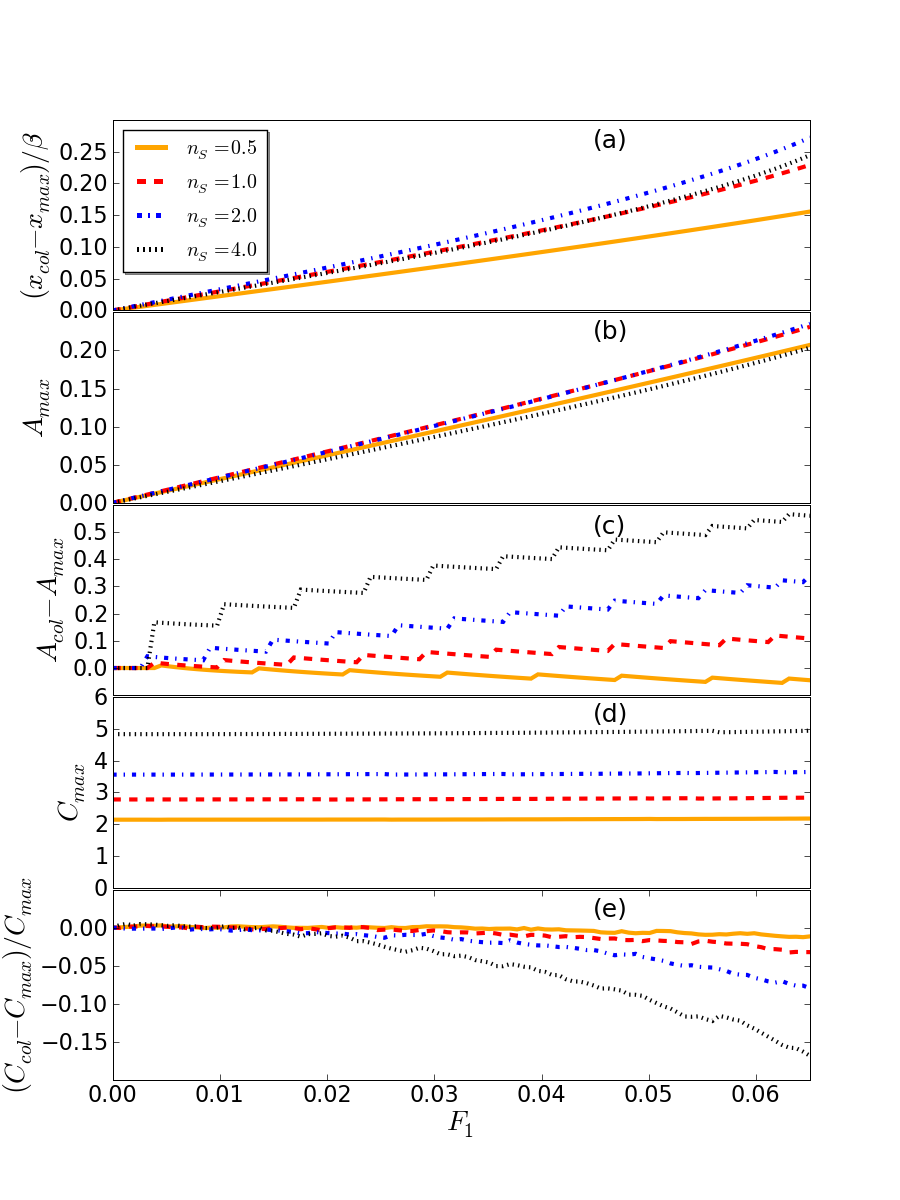}
\caption{Impact of lopsidedness on centroid (a), asymmetry with respect to maximum (b), absolute difference of asymmetries with respect to centre of light and maximum (c), concentration with respect to maximum (d), and relative difference of concentrations with respect to centre of light and maximum (e). Lopsidedness leads to a difference in maximum position and centre of light. Furthermore, lopsidedness creates asymmetry. Asymmetries evaluated with respect to the maximum or centre of light can differ substantially given that $A\in[0,2]$. The concentration evaluated at the maximum position is almost insensitive to lopsidedness. However, the concentration with respect to centre of light is strongly underestimated. All S\'ersic profiles are evaluated on a 1000$\times$1000 pixel grid using $\beta=50$. See footnote for explanation of the steps in panels (c) and (e).}
\label{fig:lopsidedness_on_x0_A_C}
\end{figure}

In Fig. \ref{fig:lopsidedness_on_x0_A_C} we investigate the impact of this type of asymmetry on the centroid, the asymmetry index and the concentration index. The first and foremost consequence is that in the presence of asymmetry the maximum position and the centre of light as given by
\begin{equation}\label{eq:def:centroid_estimation}
\hat{\vec x}_0 = \langle\vec x\rangle = \frac{\sum_n f_n \vec x_n}{\sum_n f_n} \,\textrm{,}
\end{equation}
where $\vec x_n$ and $f_n$ denote the position vector and value of pixel $n$, do not coincide anymore. Hence, we call this special type of asymmetry ``lopsidedness''. The centre of light $\vec x_\textrm{col}=\langle\vec x\rangle$ and the maximum position $\vec x_\textrm{max}$ coincide if and only if the light distribution is symmetric. As is evident from Fig. \ref{fig:lopsidedness_on_x0_A_C}, the lopsidedness is stronger for steeper profiles, where the maximum lopsidedness is $|\vec x_\textrm{col}-\vec x_\textrm{max}|/\beta\approx 0.25$. Moreover, Fig. \ref{fig:lopsidedness_on_x0_A_C} demonstrates that, especially for steep profiles, estimates of asymmetry and concentration strongly depend on the choice of centroid. Asymmetry indices estimated with respect to maximum and centre of light may differ substantially in the presence of lopsidedness considering the allowed parameter range.\footnote{The steps in panel (c) are due to the computation of $A_\textrm{col}$, since $\vec x_\textrm{col}$ is changing as $F_1$ increases. Whenever $\vec x_\textrm{col}$ enters a new pixel, the set of pixels used for computing $A_\textrm{col}$ changes. There are also steps in $C_\textrm{col}$, but they are very small.} Moreover, Fig. \ref{fig:lopsidedness_on_x0_A_C} reveals that the concentration estimated with respect to the maximum position is almost insensitive to lopsidedness, whereas the concentration estimated with respect to the centre of light can be biased low by up to 15\%. This also explains to some extent why the observed and predicted concentration indices differ in Fig. \ref{fig:C_vs_n_S_for_ZEST}, because the observed concentration indices were estimated with respect to to the centre of light rather than the maximum position \citep[cf.][]{Scarlata2007}.

We have demonstrated that the parametrisation results differ significantly depending on whether we use the centre of light or the maximum position as centroid. How do we resolve this ambiguity? And how do we get the maximum position in practice, when we suffer from pixel noise? If the parametrisation scheme was model-based, the model would define the centroid during the fit procedure -- even in the presence of pixel noise. For instance, the S\'ersic profile should use the maximum position as centroid, whereas shapelets can use both maximum position or centre of light. However, since $C$, $A$ and $M_{20}$ are not model-based, we have to resort to convention or ad-hoc solutions. In case of the asymmetry index, \citet{Conselice2000b} solved this problem by searching for the position that minimises the value of the asymmetry index, also considering resampling the image on a refined pixel grid. They were able to show that there are usually no local minima of asymmetry indices and hence that their method is stable. In case of the concentration index, using the maximum position appears to be more plausible than the centre of light, since $C_{max}$ appears to be robust against lopsidedness. Unfortunately, the concentration does not provide us with model and residuals, hence we cannot estimate the most likely maximum position in the presence of noise. However, we can apply the same ad-hoc solution that \citet{Conselice2000b} introduced for the asymmetry index, by searching the position that \textit{maximises} the concentration estimate. Nevertheless, this method increases the computational effort tremendously such that the required computation time is approximately of the same order as, e.g., fitting a shapelet model. We conclude that concentration and asymmetry estimates are neither easy to implement nor computationally faster than model-based approaches. In case of $M_{20}$, there is a theoretical preference to use the centre of light, since it minimises the total second moments.

\subsection{Example IV: Impact of lopsidedness on ellipticity estimators\label{sect:impact_A_on_eps}}

As our last example, we discuss the impact of asymmetry on estimators of ellipticity. Again, we simulate asymmetry as lopsidedness as in the previous section. We apply flexion transformations to two-dimensional S\'ersic profiles without noise. However, we do \textit{not} apply shear transformations, i.e. all profiles have no intrinsic ellipticity. From the pixellised images we then estimate the second moments of the light distribution,
\begin{equation}\label{eq:moments_Q}
Q_{ij} = \frac{\sum_n I_n (x_{n,i}-x_{0,i})(x_{n,j}-x_{0,j})}{\sum_n I_n} \;\textrm{,}
\end{equation}
where $\vec x_0$ is the point of reference, e.g. centre of light or maximum position. Using the second moments, we compute the estimator \citep[e.g.][]{Bartelmann2001}
\begin{equation}
\hat\chi = \frac{Q_{11}-Q_{22}+2i Q_{12}}{Q_{11}+Q_{22}} \;\textrm{.}
\end{equation}
This estimator is related to the axis ratio via $q=\frac{b}{a} =\sqrt{\frac{1-|\hat\chi|}{1+|\hat\chi|}}\leq 1$ and to the orientation angle $\theta$ via $\tan(2\theta)=\frac{\Im(\hat\chi)}{\Re(\hat\chi)}$. If this estimator detects any ellipticity, it will be completely artificial, i.e. it will be a bias.

\begin{figure}
\includegraphics[width=84mm]{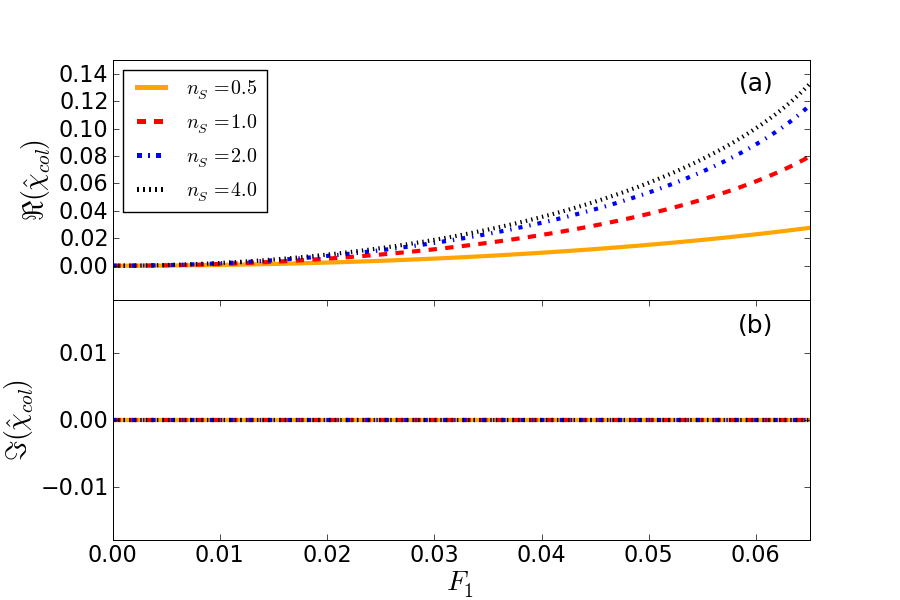}
\caption{Impact of lopsidedness on real (a) and imaginary (b) part of $\hat\chi_\textrm{col}$. Considering $0\leq|\hat\chi_\textrm{col}|<1$, the real part is strongly biased by the lopsidedness. The imaginary part is unbiased due to the geometry of $F_1$ (cf. Fig. \ref{fig:flexed_gaussians}). All S\'ersic profiles are evaluated on a 1000$\times$1000 pixel grid using $\beta=50$.}
\label{fig:lopsidedness_on_eps}
\end{figure}

Figure \ref{fig:lopsidedness_on_eps} shows results of this simulation. For perfectly symmetric profiles ($F_1=0$) the estimator indeed does not detect any ellipticity. However, if $F_1$ increases, the ellipticity estimator will be biased. The bias is stronger for steeper profiles. The maximum bias is $\Re(\hat\chi_\textrm{col})\approx 0.13$ ($b/a\approx 0.877$), which is substantial.

We conclude from this simulation that asymmetries have a potentially strong impact on ellipticity estimates, i.e. asymmetry and ellipticity are intertwined. For instance, this is relevant in case of using elliptical isophotes for estimating the concentration index.

\subsection{Reliability assessment\label{sect:reliability_assessment}}

In the previous sections we have demonstrated that some important morphological observables cannot be measured independently of one another. Given this, it cannot be guaranteed that estimates of an individual observable will result in a parametrisation which is unbiased by the other observables. As all the parametrisation schemes mentioned in Sect. 2 are derived on rather restrictive assumptions (cf. Sect. \ref{sect:assumptions}), their flexibility in describing arbitrary galaxy morphologies is therefore limited. Consequently, it cannot be expected that these schemes provide accurate descriptions of \textit{all} individual objects in a given data sample.

Can we assess the quality or reliability of the parametrisation results for \textit{individual} objects, i.e., can we detect objects where the parametrisation failed in order to sort them out?\footnote{Note that this task is completely different from testing the reliability using simulations. Such simulations allow to assess and calibrate a parametrisation scheme in general, but they do \textit{not} help in detecting parametrisation failures for \textit{individual} objects.} If we are using a model-based parametrisation scheme (e.g. shapelets or S\'ersic profiles), the residuals of the resulting best fit will provide us with an estimate of the goodness of fit. For instance, a very large value of $\chi^2$ compared to the number of degrees of freedom indicates a poor fit, i.e. we should not rely on the parametrisation of this individual object. However, if the parametrisation scheme is not model-based -- as in case of CAS, $M_{20}$ and Gini -- we have no residuals and hence we have no way of assessing the reliability for individual objects.\footnote{Note that reliability assessment and error estimation are two different things. Error estimation is possible for model-independent approaches, e.g. via bootstrapping.}

\subsection{How to disentangle observables}

As we showed above, morphological observables are intertwined and cannot be measured independently. Is there a way to get independent estimates?

Let us consider two morphological observables $A$ and $B$ (e.g. S\'ersic index and ellipticity). Intertwinement means that the joint probability of $A$ and $B$ does not factorise, i.e.
\begin{equation}
\textrm{prob}(A,B|\textrm{data}) \neq \textrm{prob}(A|\textrm{data})\,\textrm{prob}(B|\textrm{data}) \;\textrm{.}
\end{equation}
Using Bayes' theorem, we can rewrite the joint probability of $A$ and $B$ as
\begin{equation}
\textrm{prob}(A,B|\textrm{data}) = \frac{\textrm{prob}(A,B)\,\textrm{prob}(\textrm{data}|A,B)}{\textrm{prob}(\textrm{data})} \;\textrm{,}
\end{equation}
where $\textrm{prob}(A,B)$ denotes the prior probability of $A$ and $B$, $\textrm{prob}(\textrm{data}|A,B)$ is the likelihood function and $\textrm{prob}(\textrm{data})$ a normalisation factor. A model that simultaneously measures $A$ and $B$ will provide us with the likelihood function, which in case of Gaussian residuals is
\begin{equation}
\textrm{prob}(\textrm{data}|A,B) \propto e^{-\chi^2/2} \;\textrm{.}
\end{equation}
We then get independent estimates of $A$ and $B$ via marginalisation
\begin{equation}\label{eq:marginalisation_A}
\textrm{prob}(A|\textrm{data}) = \int dB\,\textrm{prob}(A,B|\textrm{data})  \;\textrm{,}
\end{equation}
\begin{equation}\label{eq:marginalisation_B}
\textrm{prob}(B|\textrm{data}) = \int dA\,\textrm{prob}(A,B|\textrm{data})  \;\textrm{.}
\end{equation}
Obviously, this only works for model-based parametrisation schemes, since otherwise we do not have residuals and cannot evaluate the likelihood function. In other words, even if we found a model-independent parametrisation scheme that accounts for all observables simultaneously, we would not know how to disentangle the estimates. In addition to reliability assessment, this is another strong argument in favour of model-based approaches.

The marginalisation integrals of Eqs. (\ref{eq:marginalisation_A}) and (\ref{eq:marginalisation_B}) are usually very hard to evaluate, unless we use Markov-Chain Monte-Carlo \citep[MCMC, e.g.][]{MacKay2008} methods. In case of MCMC methods, we get those marginalisations for free, without any further effort.


\section{Impact of PSF on the concentration index\label{sect:impact_PSF}}

In Sect. 3, we introduced the notion of intertwinement that may systematically influence morphological parameters. Another important origin of systematic effects is the point-spread function (PSF), as we illustrate in this section. The fact that parameters such as the concentration index may be influenced by the PSF is not new but has been long known. For instance, \citet{Scarlata2007} find that the PSF has a significant effect for objects with half-light radii smaller than two FWHM of the HST ACS PSF and with high Sersic index, while the effect is negligible for larger objects. In an attempt to overcome this bias, \citet{Ferreras2009} applied a correction to the measured concentration parameter, based on the half-light radius. The aim of this section is to reassess the impact of the PSF on estimates of the concentration index.

\subsection{Forward vs. backward PSF modelling}

In case of model-based parametrisation schemes it is standard practice to account for the PSF by forward modelling, i.e. to fit a convolved model to the convolved data. In case of parametrisation schemes that are not model-based this is impossible and we have to resort to backward PSF modelling, i.e. we deconvolve the data before the actual parametrisation is done. However, deconvolution in the presence of pixel noise is numerically unstable, so forward PSF modelling is to be favoured if possible. This is another practical disadvantage of parametrisation schemes that are not model-based, because they need to perform either an unstable backward modelling or they need to invoke another ad-hoc correction calibrated in simulations. Such simulation-based calibrations introduce a further assumption into the parametrisation process. Model-based schemes are much more rigorous in this respect, since they allow for a mathematically well-defined PSF treatment that does not introduce any further assumption.

\subsection{Impact on concentration\label{sect:impact_PSF_on_C}}

In case of the ZEST, \citet{Sargent2007} accounted for the PSF by forward modelling when estimating the S\'ersic index, while \citet{Scarlata2007} neglected the PSF when estimating the concentration index. The fact that the results shown in Fig. \ref{fig:C_vs_n_S_for_ZEST} are in agreement with theoretical predictions suggests that in the case of the COSMOS data the PSF can indeed be neglected for the concentration index. Therefore, the theoretical prediction supports the claim by \citet{Scarlata2007}. Nevertheless, this single example should not mislead us to generalise this conclusion. It is \textit{not} guaranteed that the PSF will have no impact on the concentration index for data sets other than COSMOS that exhibit different signal-to-noise, PSF, and resolution.

\begin{figure}
\includegraphics[width=84mm]{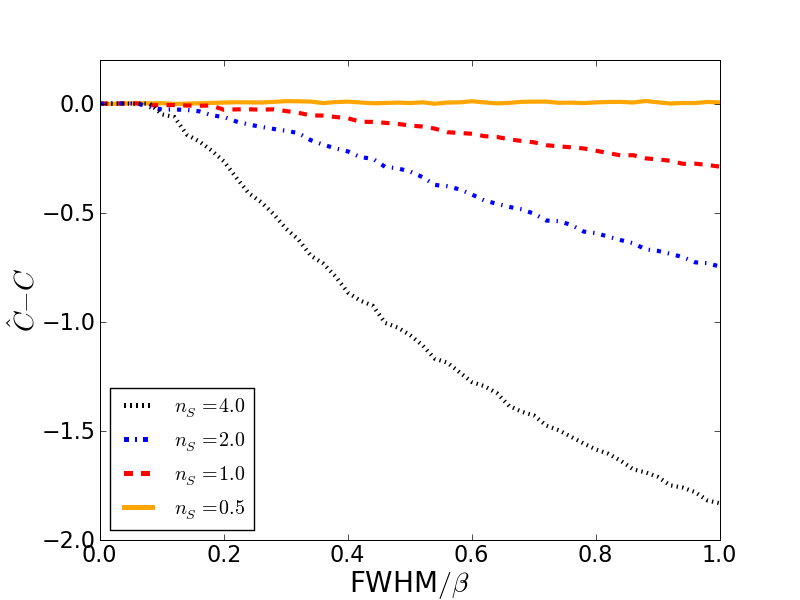}
\caption{Impact of PSF on misestimation $\hat C-C$ of concentration index for different PSF sizes and S\'ersic profiles. All S\'ersic profiles are evaluated on a 1000$\times$1000 pixel grid using $\beta=50$ and $b_n=2 n_S - 1/3$. With increasing PSF size with respect to the object size the concentration index estimated from the convolved image is more and more underestimated.}
\label{fig:impactPSFonC}
\end{figure}

In order to test the impact of the PSF on the concentration index, we generate two-dimensional S\'ersic profiles with $n_S=0.5,1,2,4$ and convolve these profiles with a Gaussian kernel of increasing FWHM.\footnote{We are aware that the COSMOS PSF is not a Gaussian. This test is meant to demonstrate the principle of this effect.} We expect that the concentration indices of very steep S\'ersic profiles are severly underestimated, since the PSF washes out the sharp peak. For lower S\'ersic indices this effect becomes smaller. For $n_S=0.5$ the concentration should not be affected at all, since convolution of a Gaussian with a Gaussian yields a Gaussian, i.e. the steepness of the profile does not change. Figure \ref{fig:impactPSFonC} confirms our expectation. If we ignore the PSF, we can significantly underestimate the concentration index.

We conclude from this test that although the PSF is indeed negligible in case of the ZEST, this cannot be generalised to other data sets. Consequently, a PSF treatment is always necessary at least when using the concentration index. In particular concerning ground-based telescopes, the PSF is usually \textit{not} small compared to the peak exhibited by highly concentrated objects.


\section{Parametrisation \& classification\label{sect:para_&_classification}}

We now discuss the parametrisation of galaxy morphologies in the context of classification. First, we show that if we do not account for all morphological observables simultaneously, the effects discussed in the previous sections can dilute discriminative information. Second, we show that all parametrisation schemes discussed here form nonlinear or even discontinuous parameter spaces. Third, we comment on the problem of high-dimensional parameter spaces.

\subsection{Loss of discriminative information\label{sect:loss_discri_info}}

The conclusion from our investigation of the intertwinement was: If a parametrisation scheme does not account for all morphological observables simultaneously, the results will be systematically altered, i.e. biased. How does this influence classification results? For a large sample of objects, the origins of these systematic effects have random strength. Consequently, we have to expect an increase in the scatter of the resulting parameters. The sample distributions of the parameters will be broadened due to the additional scatter, i.e. peaks in the distributions are reduced and troughs between different peaks are washed out. In other words, we are loosing discriminative information.

We now demonstrate this broadening of parameter distributions: We generate samples of two-dimensional S\'ersic profiles with fixed S\'ersic indices of $n_S=1,2,3,4$. We then add a random ellipticity and a random lopsidedness via the flexion transformation of Eq. (\ref{eq:flexion_trafo}). The flexion parameter $F_1$ is drawn from a uniform distribution on the interval $[-0.065,0.065]$. The ellipticity is drawn from the joint distribution of S\'ersic indices and axis ratios of 2,000 COSMOS galaxies randomly drawn from the Zurich Structure \& Morphology catalogue. We then sample the S\'ersic profiles on a 1,000$\times$1,000 pixel grid using a scale radius of $\beta=50$. We convolve the resulting image with a Gaussian PSF of FWHM$=37.5$ chosen such that the effects of Fig. \ref{fig:impactPSFonC} are present but moderate. There is no pixel noise in this simulation. From the pixellised image we then estimate the concentration with respect to the maximum position and the centre of light, since S\'ersic index and concentration are two different estimators for the same morphological feature. Concentration estimates also take into account elliptical isophotes, where the ellipticity is estimated via Eq. (\ref{eq:moments_Q}) with respect to the maximum position and the centre of light, respectively.

\begin{figure}
      \includegraphics[width=84mm]{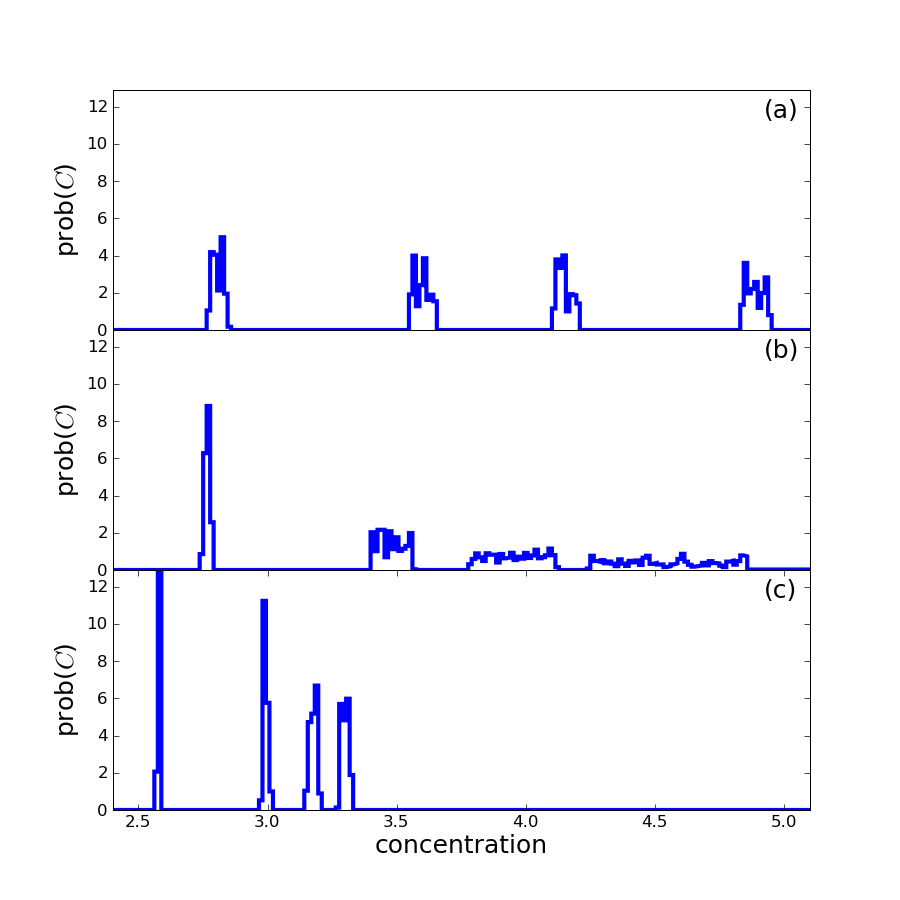}
\caption{Normalised sample distributions of concentration indices estimated with respect to (a) the maximum position of unconvolved image, (b) the centre of light of unconvolved image, and (c) the centre of light of convolved images. The modes in the distributions correspond to samples of 10,000 profiles each with fixed S\'ersic indices of exactly $n_S=1,2,3,4$ (from left to right). The finite widths of all modes in all distributions indicate the loss of discriminative information. This is particularly evident in panel (b), where the modes of very compact objects are substantially broadened. All S\'ersic profiles were evaluated on a 1000$\times$1000 pixel grid using a scale radius of $\beta=50$. The Gaussian convolution kernel for panel (c) was evaluate on the same pixel grid with FWHM$=37.5$.}
\label{fig:broadened_distribution_C}
\end{figure}

Figure \ref{fig:broadened_distribution_C} shows the results of this simulation. The distributions of concentration indices have a finite width, in contrast to the distribution of the S\'ersic indices, which are infinitely thin $\delta$-peaks. Consequently, we are indeed loosing discriminative information. In reality this loss may be even more severe, since the distribution of S\'ersic indices has itself a finite width. Moreover, Fig. \ref{fig:broadened_distribution_C} reveals that the loss of discriminative information is stronger for the concentration index evaluated at the centre of light. Especially for large S\'ersic indices the peaks are lowered and broadened. This is a strong argument to evaluate the concentration at the maximum position (if it were accessible), since we conserve more discriminative information. In the presence of an unconsidered PSF, the parameter space is substantially biased. This has the advantage of reducing the width of the distributions, but it also shifts the different modes closer together. If the distribution of S\'ersic indices had a finite width, this would wash out the troughs separating the peaks.

This simulation demonstrates that an incautious use of the concentration index (ignoring asymmetries and the PSF) can lead to a substantial loss of discriminative information. In practice, this loss causes sample distributions of the concentration index to be of low modality, despite the diversity of the galaxy population -- a problem already mentioned by \citet{Faber2007}. Consequently, the concentration index can only provide a lower bound on the number of classes in a given data sample. If the sample distribution of the concentration is unimodel, this does \textit{not} imply that all objects are of the same type. The loss of discriminative information implies that the mapping $\mathcal F_2^{-1}$ from Sect. \ref{sect:trinity} does not always exist for the concentration index, i.e. drawing inference is a very difficult task.

\subsection{Nonlinear \& discontinuous parameter spaces}

This section highlights an additional problem, which is independent of the previous considerations. It is based on the fact that all parametrisation schemes discussed here are nonlinear in the data. As a direct consequence of this, the resulting parameter spaces form nonlinear spaces, too. If the parameter space is nonlinear, the distance metric will be nonlinear, too. Although this fact may be known, it is typically ignored in practice. Usually, the Euclidean metric is employed whenever a distance-based algorithm is used, e.g., a principal components analysis \citep{Scarlata2007} or classification algorithms \citep[e.g.][]{Gauci2010}. The crucial question is: Does ignoring the nonlinearity and employing the Euclidean distance leads us to misestimate the true distances between galaxy morphologies in the parameter space? If so, galaxies will seem more similar or less similar than they actually are and hence distance-based classification algorithms may face serious problems. There are only few classification algorithms that do not rely on distances \citep[e.g.][]{Fraix-Burnet2009}.

\subsubsection{Nonlinearity\label{sect:nonlinearity_of_schemes}}

Let us consider a parametrisation $P(I)$ of an image $I$. This parametrisation is said to be \textit{linear} in the image data, if
\begin{equation}
P(\alpha\,I_A + \beta\,I_B) = \alpha\,P(I_A) + \beta\,P(I_B)
\end{equation}
for any two images $I_A$ and $I_B$ and any real-valued $\alpha$ and $\beta$. Otherwise $P$ is nonlinear.

We begin by considering CAS (Eqs. (\ref{eq:def:concentration})--(\ref{eq:def:clumpiness})). Apart from the obvious nonlinearities in $C$ due to the logarithm and the ratio of radii, the computation of the radii containing 20\% and 80\% of the total flux itself is highly nonlinear. The nonlinearities in $A$ and $S$ are caused by the fractions and absolute values in the numerators. Gini (Eq. (\ref{eq:def:gini})) and $M_{20}$ (Eq. (\ref{eq:def:M20})) are both nonlinear in the data, too. For both of them the major nonlinearity is hidden in the sorting of the pixel values. The S\'ersic model given by Eq. (\ref{eq:def:Sersic_model}) contains the S\'ersic index and the scale radius as nonlinear parameters.

The nonlinearity of (spherically symmetric) shapelets is due to the scale radius $\beta$ and the centroid $\vec x_0$. Both enter the basis functions nonlinearly, as is evident from Eq. (\ref{eq:def:shapelets}). The nonlinearity of shapelets has been investigated in detail by \citet{Melchior2007}, so we do not need to elaborate on this here. In case of s\'ersiclets, the S\'ersic index is another nonlinear model parameter in addition to the scale radius.

\subsubsection{Demonstration of nonlinearity of $C$, $A$ \& Gini}

\begin{figure}
\begin{center}
      \includegraphics[width=3.5cm]{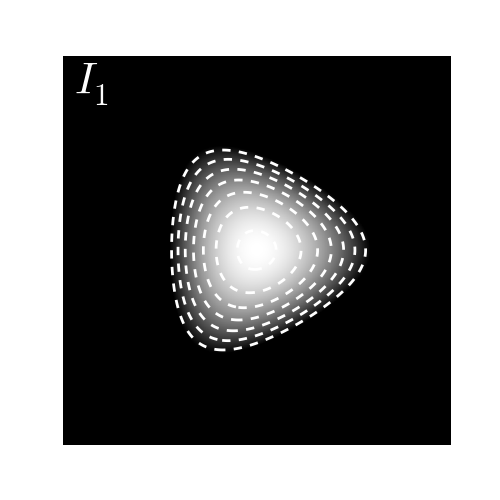}
      \includegraphics[width=3.5cm]{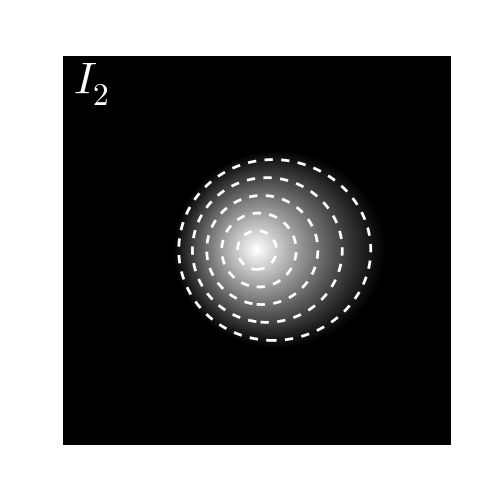}
      \includegraphics[width=3.5cm]{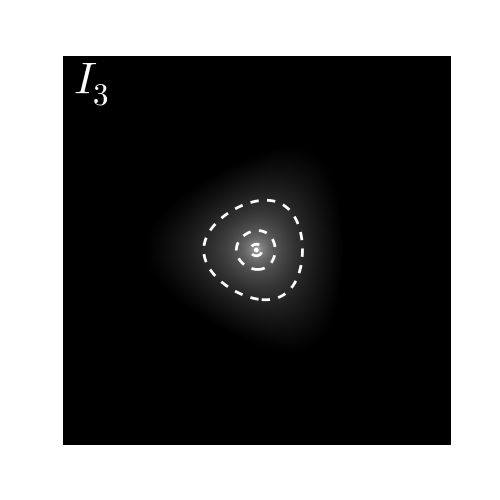}
\end{center}
\caption{Two-dimensional profiles with different asymmetries used for demonstration of nonlinearity. All objects are evaluated on a 1,000$\times$1,000 pixel grid with scale radius $\beta=50$. No intrinsic ellipticity was applied. All maximum positions are identical. Profile $I_1$ (top left) has flexion $G_1=0.1$ and $n_S=0.5$. Profile $I_2$ (top right) has flexion $F_1=0.05$ and $n_S=1$. Profile $I_3$ (bottom) has flexion $G_1=-0.1$ and $n_S=4$.}
\label{fig:galaxies_ABC}
\end{figure}

As emphasised above, CAS, Gini, $M_{20}$ and the S\'ersic index are nonlinear in the data. The crucial question is: Is the nonlinearity severe or can we assume local flatness in the parameter space and use the Euclidean metric as an approximation? In order to answer this question, we now show a demonstration using three S\'ersic profiles with different S\'ersic indices and different flexion values as shown in Fig. \ref{fig:galaxies_ABC}. There is no pixel noise in this simulation. We perform a linear transformation in the image space such that two images $I_A$ and $I_B$ linearly transform into each other, i.e.
\begin{equation}\label{eq:linear_transformation}
I(\alpha) = (1-\alpha)I_A + \alpha I_B \;\textrm{,}
\end{equation}
where $\alpha\in[0,1]$ parametrises this linear transformation. In reality, the superpositions of this linear transformation may not represent viable galaxy morphologies, e.g. $\alpha=0.5$ for $I_1\leftrightarrow I_3$. A proper trajectory should be a geodesic on the submanifold of viable morphologies. If this submanifold is linear, the trajectory defined by Eq. (\ref{eq:linear_transformation}) will pass through viable morphologies only. If it is nonlinear, it will add additional nonlinearity to this test. This means that even though Eq. (\ref{eq:linear_transformation}) passes through unrealistic morphologies in this setup, it provides a lower limit to the nonlinearity. For 100 equidistant values of $\alpha\in[0,1]$ we evaluate the mixed image $I(\alpha)$ in pixel space and then estimate the concentration and asymmetry with respect to the maximum position. We also estimate the Gini coefficient. 

\begin{figure*}
      \includegraphics[width=5.5cm]{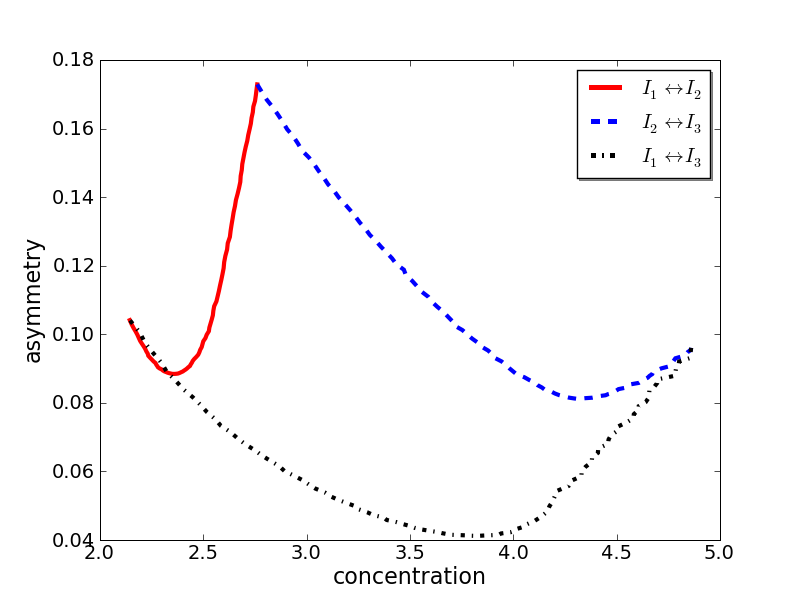}
      \includegraphics[width=5.5cm]{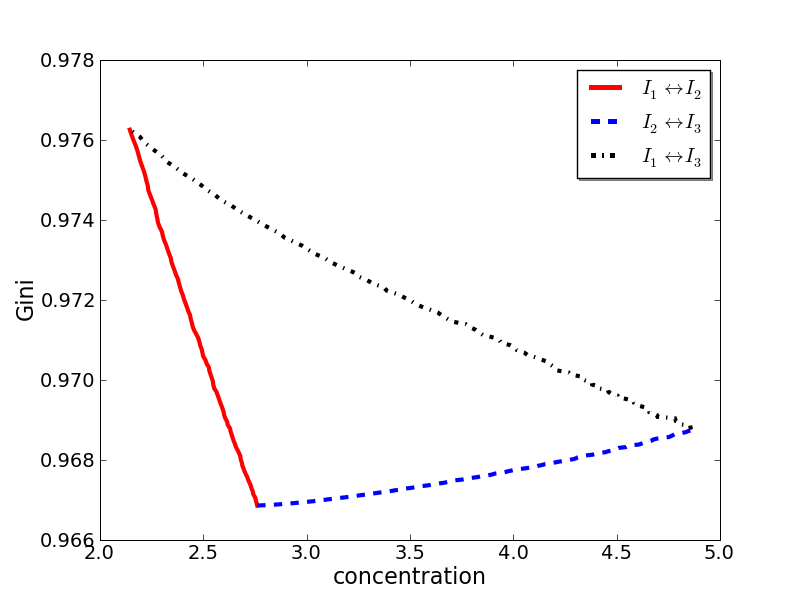}
      \includegraphics[width=5.5cm]{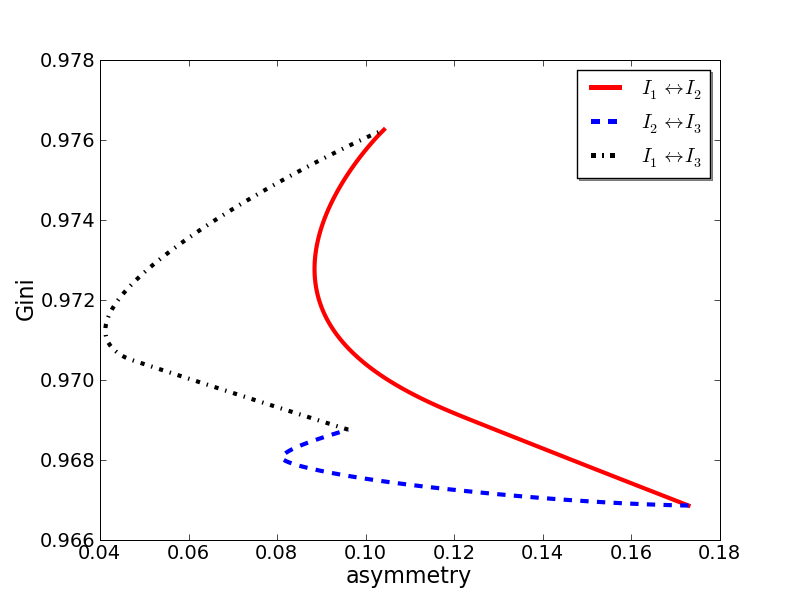}
\caption{Trajectories in CA-Gini subspaces revealing substantial nonlinearities. Left panel: Trajectories in CA space. Centre panel: Trajectories in C-Gini space. Right panel: Trajectories in Gini-A space. In this simulation the nonlinearity is induced by the different lopsidedness of all objects (cf. Fig. \ref{fig:galaxies_ABC}). The asymmetry is evaluated with respect to the maximum position, whereas the concentration is evaluated with respect to the centre of light.}
\label{fig:trajectories_CAG}
\end{figure*}

Figure \ref{fig:trajectories_CAG} shows the trajectories in the subspaces of $C$, $A$ and Gini. Example objects $I_1$ and $I_2$ have very similar S\'ersic indices and flexion parameters, hence their transition produces trajectories that are only moderately nonlinear. However, example object $I_3$ is very different from $I_1$ and $I_2$ and thus its transitions produce trajectories that exhibit substantial nonlinearities. Note that the nonlinearities in Fig. \ref{fig:trajectories_CAG} are primarily induced by the lopsidedness via the asymmetry parameter, as is evident from the centre panel where $A$ is not shown and virtually all nonlinearity is gone.

We conclude from this simulation that for galaxy morphologies exhibiting realistic asymmetries the Euclidean distance is a very poor approximation to distances in parameter space. Consequently, any algorithm based on Euclidean distances would severely underestimate the true distances, i.e. objects would appear more similar than they actually are. This may be an explanation why the drop in the spectrum of eigenvalues of the principal components analysis of \citet{Scarlata2007} -- which justifies the reduction of dimensionality -- is not very decisive. It may also partially account for the difficulty of recovering visual classifications using automated algorithms \citep[see e.g.][]{Gauci2010}. This is no particular drawback of $C$, $A$ and Gini, but applies to all other parametrisation schemes discussed here. It is highly questionable whether a ``calibration'' of the Euclidean distance in order to account for the nonlinearity is possible. The reason for this is that, due to nonlinearity, the distance is an unknown function of the positions of both objects in parameter space, i.e. the distance depends on the morphology. One possible solution is to try to estimate the true distance via a linear transformation as given by Eq. (\ref{eq:linear_transformation}), although that is computationally very expensive. Another option is to employ a method called ``diffusion distance'' \citep{Richards2009} in order to estimate the true nonlinear distances.

\subsubsection{Discontinuity of spaces formed by $C$ and $M_{20}$\label{sect:discontinuity_C_M20}}

\begin{figure}
      \includegraphics[width=84mm]{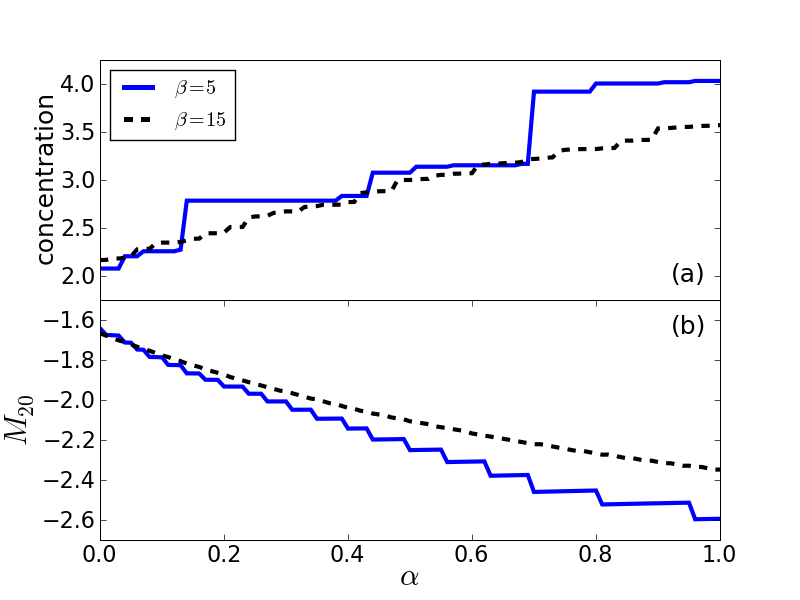}
\caption{Discontinuity of concentration (a) and $M_{20}$ (b). For poor sampling (small $\beta$), concentration index and $M_{20}$ exhibit substantial discontinuities. For better sampling (larger $\beta$) the discontinuities decrease. The transition was between two S\'ersic profiles with $n_S=0.5$ and $n_S=2.0$ and no intrinsic ellipcitity or lopsidedness. The scale radii were $\beta=5$ (blue lines) and $\beta=15$ (red lines), respectively. The profiles were evaluated on a 300$\times$300 pixel grid.}
\label{fig:discontinuity_C_M20}
\end{figure}

In Fig. \ref{fig:discontinuity_C_M20} we investigate the behaviour of concentration and $M_{20}$ under a linear transformation between two S\'ersic profiles. $C$ and $M_{20}$ exhibit substantial discontinuities due to pixellation effects. These effects increase for decreasing resolution (i.e. decreasing $\beta$ in Fig. \ref{fig:discontinuity_C_M20}).

In the case of $C$, the discontinuities occur because the radii containing 20\% and 80\% of the total image flux can only change in discrete steps. With increasing resolution, the pixel size decreases and the discontinuities of $R_{20}$ and $R_{80}$ become smaller (cf. panels (a) and (b) in Fig. \ref{fig:discontinuity_C_M20}). Hence, this is not a problem for well resolved galaxies as in Fig. \ref{fig:trajectories_CAG}. However, it is a problem for poorly sampled galaxies. In this case, we can overcome this problem by interpolating the pixellised image and integrating numerically. Unfortunately, this would drastically increase the computational effort. In fact, the discontinuity of the concentration index has already been observed by \citet{Lotz2006}.

In the case of $M_{20}$, the origin of the discontinuity is the sum over the second-order moments in the numerator of Eq. (\ref{eq:def:M20}), which stops as soon as 20\% of the total flux are reached. This threshold is the problem, as it causes the set of pixels fulfilling this criterion to change abruptly during the linear transformation. Again, the discontinuities of $M_{20}$ decrease with increasing resolution. However, for poorly sampled galaxies we cannot overcome these discontinuities by interpolation, since the definition of $M_{20}$ only makes sense for pixellised images.

A parametrisation scheme forming discontinuous parameter spaces is problematic, because it is not guaranteed that objects with similar morphologies end up in neighbouring regions of the parameter space. This implies that distances in the space formed e.g. by $M_{20}$ do not necessarily correlate with the similarity of galaxy morphologies. \textit{We need similar morphologies to have smaller distances than dissimilar morphologies}, but this is not guaranteed for $C$ and $M_{20}$ if the resolution is poor. Figure \ref{fig:discontinuity_C_M20} suggests that such discontinuities become important when galaxies are smaller than 10 pixels in radius, maybe even earlier depending on the precise morphology. In this case, we even cannot rely on hard-cut classifications and it is questionable whether meaningful classification based on distances is possible at all.

\subsection{High-dimensional parameter spaces}

Concerning classification, the current paradigm appears to favour low-dimensional parameter spaces \citep[e.g.][]{Scarlata2007} that simplify the analysis or even allow a visual representation. However, we have to keep in mind that a high-dimensional parameter space may be necessary in order to differentiate between different groups of galaxy morphologies. There is no physical reason to expect that a two- or even three-dimensional parameter space should be able to host such groups without washing out their differences. This solely depends on the complexity of the physics governing galaxy morphologies.

In particular, basis-function expansions typically form parameter spaces of high dimensionality. For instance, the morphological parameter space used by \citet{Kelly2005} had 455 dimensions. Apart from problems with visualisation, we suffer from what is commonly called the {\it curse of dimensionality} \citep{Bellman1961}: The hypervolume of a (parameter) space grows exponentially with its number of dimensions.\footnote{Consider a hypercube  of edge length $L$ in $d$ dimensions. Its hypervolume $L^d$ grows exponentially with $d$.} Consequently, the density of data points in this parameter space is suppressed exponentially. Therefore, it is impossible to reliably model a data distribution in a parameter space of several hundred dimensions, no matter how much data is available. Nevertheless, it is preferable to employ a parametrisation scheme that produces a high-dimensional parameter space. Loosely speaking, it is better to start with too much information than with too little. We can overcome the curse of dimensionality, if we compress the parameter space, i.e. if we reduce its number of dimensions by identifying and discarding unimportant or redundant information. For instance, \citet{Kelly2004,Kelly2005} applied a principal component analysis in order to reduce the dimensionality of their parameter space.\footnote{Unfortunately, a principal component analysis (PCA) is a risky and often inappropriate tool in the context of classification. The reason is that PCA diagonalises the sample covariance matrix, i.e., it assumes that the whole data sample comes from a \textit{single} Gaussian distribution. This assumption obviously jars with the goal of assigning objects to \textit{different} classes.} An alternative approach to overcome the curse of dimensionality is to employ a kernel approach by describing the data using a similarity measure. We demonstrated in \citet{Andrae2010a} that this yields excellent results, e.g., allowing us to classify 84 galaxies populating a 153-dimensional parameter space into three classes.

\section{Summary \& conclusions\label{sect:summary}}

In this paper we have described and compared two different approaches to the parametrisation of galaxy morphologies: First, model-independent schemes -- CAS, Gini and $M_{20}$. Second, model-based schemes -- S\'ersic profiles and basis functions.

Our most important result is that morphological features (steepness of light profile, ellipticity, asymmetry, substructures, etc.) are intertwined and (at least some) cannot be estimated independently without introducing potentially serious biases. This intertwinement stems from the violation of one or more assumptions invoked by the parametrisation schemes. We emphasise that combining separate estimates of individual observables does \textit{not} overcome the intertwinement. For instances, combining an ellipticity estimate and the fit of a circular S\'ersic profile does not give the same result as fitting an elliptical S\'ersic profile. No parametrisation scheme discussed in this article accounts for all these observables simultaneously, i.e., their usage will inevitably cause problems when trying to parametrise large samples of galaxies that exhibit a huge variety of morphologies.

In the context of classification of galaxy morphologies, which is an important application, we have the following results:
\begin{itemize}
\item The intertwinement can wash out discriminative information in the context of classification.
\item All parametrisation schemes form nonlinear parameter spaces with a potentially highly nonlinear and unknown metric. Distance-based classification algorithms that employ the Euclidean distance measure therefore suffer from a loss of discriminative information.
\item For poorly resolved galaxies (object radius smaller than $\approx$10 pixels), concentration and $M_{20}$ form discontinuous parameter spaces that do not conserve neighbourhood relations of morphologies and may therefore fool classification algorithms.
\end{itemize}
Due to the complexity of a nonlinear metric, it appears unlikely that calibrating results obtained from Euclidean distance is possible. As we cannot expect to find a parametrisation scheme that is linear in the data, a more promising approach is to estimate the nonlinear metric, e.g. via diffusion distances \citep{Richards2009}, or to use a classification algorithm that is not distance-based. An example for such an algorithm can be found in \citet{Fraix-Burnet2009}.

\subsection{Arguments in favour of model-based approaches}

In this paper we also collected arguments in favour of model-based approaches:
\begin{itemize}
\item A (compact) model defines the term ``centroid'', i.e. whether we have to use the centre of light or the maximum position.
\item A model allows us to disentangle observables by marginalising the joint posterior distribution of all observables.
\item A model allows us to assess reliability by providing residuals.
\item A model allows forward PSF modelling, which is more stable than backward modelling in the presence of pixel noise.
\end{itemize}
Each of these arguments by itself disfavours model-independent approaches. Therefore, we conclude that schemes such as CAS, Gini and $M_{20}$ are problematic for three reasons:
\begin{enumerate}
\item They try to measure morphological features independently ignoring their intertwinement (e.g. concentration does not account for asymmetry and vice versa).
\item They do not provide residuals, i.e. we can neither assess reliability (to sort out failures for \textit{individual} objects) nor marginalise.
\item They do not allow forward PSF modelling, i.e. we may suffer from the instability of backward modelling, or, we need to introduce further assumptions via calibrations.
\end{enumerate}
Moreover, we have seen that robust implementations of CAS and $M_{20}$ are neither easy nor computationally fast, since we have to consider centroid misestimations and -- in the case of the concentration index -- interpolation.

We conclude that model-based parametrisation schemes are clearly superior. They provide reliable parametrisation schemes in \textit{all} regimes of signal-to-noise ratios and resolutions. For low signal-to-noise ratios and low resolution the S\'ersic profile allows excellent parametrisations \citep[e.g.][]{Sargent2007}. In the limit of high signal-to-noise ratios and high resolutions the method of shapelets is flexible enough to provide excellent model reconstructions \citep[e.g.][]{Andrae2010a}. With the advent of s\'ersiclets there will be another set of basis functions that is designed to provide even better parametrisations than shapelets \citep{Andrae2010c}.

\subsection{Trade-offs}

Throughout this work we were facing two important trade-offs when comparing different parametrisation schemes for arbitrary galaxy morphologies, namely
\begin{enumerate}
\item simplicity vs. reliability and
\item interpretation vs. flexibility.
\end{enumerate}

The first trade-off -- simplicity vs. reliability -- is obvious. When dealing with large data samples, we have to find a parametrisation scheme that is not too expensive from a computational point of view. Apart from computational aspects, we also favour simple solutions in general (Occam's razor). However, we have to beware of \textit{over}simplification which inevitably leads to unreliable results. The borderline between reasonable simplification and oversimplification should be defined by the data only and \textit{not} by the researcher.

The second trade-off -- interpretation vs. flexibility -- is at the heart of this article. We have seen that parametrisation schemes that easily offer interpretation often lack flexibility (e.g. CAS), whereas other schemes (e.g. shapelets) excell in flexibility but lack interpretation. This is still an open issue and more work is needed on the interpretation of basis-function expansions.

We should also add that there is actually \textit{no} trade-off concerning computational feasibility. The parametrisation of samples of galaxies is trivial to parallelise, i.e. it can be done on numerous computers simultaneously.

\subsection{Recommendations and outlook}

We do \textit{not} conclude that CAS, Gini and $M_{20}$ should not be used anymore. According to their assumptions as given in Sect. \ref{sect:assumptions}, these parametrisation schemes are highly specialised on certain morphologies and their usage should be safe, if it is ensured that the sample of interest only contains galaxies of this special type. However, this obvious lack of flexibility renders these approaches inappropriate for general samples. Our most important recommendations for using CAS, Gini and $M_{20}$ are as follows:
\begin{itemize}
\item A PSF treatment is necessary at least in case of the concentration index.
\item Beware of undersampling effects in case of concentration index and $M_{20}$. Discontinuities can appear for objects of up to 10 pixels in radius.
\item Beware of the centroid ambiguity: Even for galaxies with realistic asymmetries the centre of light and maximum position do not coincide. In case of the concentration index, we recommend to fit for the centroid by maximising $C$, similar to the method of \citet{Conselice2000b}.
\end{itemize}
Concerning the concentration index, we also recommend to use it only in the regime of intermediate signal-to-noise ratios and resolutions. The reasons is that its assumptions (Sect. \ref{sect:assumptions}) are almost identical to the assumptions of a S\'ersic profile. As a rule of thumb we can say that the concentration index is reliable whenever the S\'ersic profile is a good description, and vice versa.

Currently the most reliable parametrisation scheme is the two-dimensional S\'ersic profile enhanced by ellipticity, since it accounts for the steepness of the light profile and for ellipticity. These are definitely the two most important morphological observables. In the presence of asymmetries we recommend defining the centroid by fitting for the maximum position of the profile rather than fixing it to the centre of light. However, the S\'ersic profile does not account for asymmetry or substructures and is thus of limited usefulness for samples containing highly irregular galaxies and in the regime of high signal-to-noise ratios and high resolutions. Moreover, we have shown that the scale radius of the S\'ersic profile is difficult to interpret. In particular we have argued that the scale radii of profiles of different S\'ersic indices \textit{cannot} be compared directly. We also demonstrated that a redefinition of the S\'ersic model may simplify the fitting procedure and provide more robust parameter estimates.

Our main conclusion is: None of the existing parametrisation schemes is applicable to the task of parametrising arbitrary galaxy morphologies that occur in large samples, since they all have their drawbacks. Therefore, we need a new parametrisation scheme. Our recommendations for its design are as follows:
\begin{enumerate}
\item It should be model-based.
\item It should estimate all relevant morphological features simultaneously.
\item It should provide excellent model reconstructions of galaxies in the regime of high signal-to-noise ratios and high resolutions.
\item It should form a metric parameter space such that it is possible to estimate distances for classification purposes.
\end{enumerate}
One possible solution is to modify e.g.\ the S\'ersic profile in order to account for asymmetries and substructures \citep[Galfit 3,][]{Peng2010}. In our opinion basis functions are also promising candidates to describe arbitrary morphologies, since they are highly flexible. However, current sets of basis functions still lack direct physical interpretation. Currently, we reinvestigate the method of s\'ersiclets which appears to be the most promising approach given the considerations of this paper.

\section*{acknowledgements}

RA thanks Eric Bell for discussions that initialised this work. RA also thanks Matthias Bartelmann, Thorsten Lisker, Aday Robaina Rapisarda, Mark Sargent, Paraskevi ``Vivi'' Tsalmantza, Glenn van de Ven, and Katherine Inskip for helpful comments on the content of this paper. Furthermore, we thank Claudia Scarlata for pointing out a mistake in an earlier version of this manuscript. RA is funded by a Klaus-Tschira scholarship. KJ is supported by the Emmy-Noether-programme of the DFG. PM is supported by the DFG Priority Programme 1177.

\bibliographystyle{mnras}

\def\physrep{Phys. Rep.}%
\def\apjs{ApJS}%
\def\apjl{ApJ}%
\def\apj{ApJ}%
\def\aj{AJ}%
\def\aap{A\&A}%
\def\aaps{A\&AS}%
\def\mnras{MNRAS}%
\bibliography{bibliography}

\appendix
\section{Shear and flexion transformation}
\label{app:shear_flexion_trafo}

We now briefly resume the shear and flexion transformation we are using to simulate ellipticity and lopsidedness -- the latter being a special kind of asymmetry.

Given the complex ellipticity, $\epsilon = \epsilon_1 + i\,\epsilon_2$, with axis ratio $q=\frac{b}{a}=\frac{1-|\epsilon|}{1+|\epsilon|}$ and orientation angle $\theta=\frac{1}{2}\arctan(\frac{\epsilon_2}{\epsilon_1})$, the ``sheared'' coordinates, $(x_1^\prime,x_2^\prime)$, are given by
\begin{equation}
\left(\begin{array}{c} x_1^\prime \\ x_2^\prime \end{array}\right)
= \left(\begin{array}{cc} 1-\epsilon_1 & -\epsilon_2 \\ -\epsilon_2 & 1+\epsilon_1 \end{array}\right) \cdot
\left(\begin{array}{c} x_1 \\ x_2 \end{array}\right) \,\textrm{.}
\end{equation}
For given pixel coordinates $(x_1,x_2)$, we then evaluate the model at $(x_1^\prime,x_2^\prime)$.

The flexion transformation \citep{Goldberg2005} is parametrised by the first flexion
\begin{equation}
F = F_1 + i F_2
\end{equation}
and the second flexion
\begin{equation}
G = G_1 + i G_2 \;\textrm{.}
\end{equation}
Given these parameters, we compute the derivatives of the gravitational shear $\boldsymbol\gamma=(\gamma_1,\gamma_2)$,
\begin{equation}
\gamma_{1,1} = \frac{1}{2}(F_1 + G_1)
\end{equation}
\begin{equation}
\gamma_{2,2} = \frac{1}{2}(F_1 - G_1)
\end{equation}
\begin{equation}
\gamma_{1,2} = \frac{1}{2}(G_2 - F_2)
\end{equation}
\begin{equation}
\gamma_{2,1} = \frac{1}{2}(G_2 + F_2) \;\textrm{.}
\end{equation}
Based on these derivatives, we compute the two matrices
\begin{equation}
D_{ij1} = \left(\begin{array}{cc}
-2\gamma_{1,1} - \gamma_{2,2}  &  -\gamma_{2,1} \\
-\gamma_{2,1}  &  -\gamma_{2,2}
\end{array}\right)
\end{equation}
and
\begin{equation}
D_{ij2} = \left(\begin{array}{cc}
-\gamma_{2,1}  &  -\gamma_{2,2} \\
-\gamma_{2,2}  &  2\gamma_{1,2} - \gamma_{2,1}
\end{array}\right) \;\textrm{.}
\end{equation}
Using these matrices, we do not evaluate a flexed S\'ersic profile at position $\vec x=(x_1,x_2)$, but rather at position
\begin{equation}\label{eq:flexion_trafo}
x_i^\prime = x_i + \frac{1}{2}D_{ijk}x_j x_k \;\textrm{.}
\end{equation}
The scaling of the coordinates by the scale radius $\beta$ of the S\'ersic profile is applied \textit{prior} to this flexion transformation.

\bsp

\label{lastpage}

\end{document}